\documentclass[10pt,journal,compsoc]{IEEEtran}
\ifCLASSOPTIONcompsoc
  \usepackage[nocompress]{cite}
\else
  \usepackage{cite}
\fi
\ifCLASSINFOpdf
\else
\fi
\usepackage{amsmath,amssymb,amsfonts}
\usepackage{algpseudocode}
\usepackage[linesnumbered,ruled,vlined]{algorithm2e}
\usepackage{pifont}
\usepackage{graphicx}
\usepackage{textcomp}
\usepackage{xcolor}
\usepackage{pgf}
\usepackage{tikz}
\usepackage{color}
\usepackage{stfloats}
\usepackage{multirow}
\usepackage{diagbox}
\usepackage[hyphens]{url}
\usepackage{soul}
\usepackage{color}
\usepackage{multirow}
\usepackage{makecell}
\usepackage{xcolor}
\usepackage{amssymb}
\usepackage{tikz}
\usetikzlibrary{calc}
\usepackage[bookmarks=true,breaklinks=true,colorlinks,linkcolor=blue,citecolor=blue,urlcolor=black]{hyperref}

\usepackage{booktabs}
\usepackage{comment}
\usepackage{ulem}

\newcommand{\squishlist}{
   \begin{list}{$\bullet$}
    { \setlength{\itemsep}{0pt}      \setlength{\parsep}{0pt}
      \setlength{\topsep}{3pt}       \setlength{\partopsep}{0pt}
      \setlength{\listparindent}{-2pt}
      \setlength{\itemindent}{-5pt}
      \setlength{\leftmargin}{1em} \setlength{\labelwidth}{0em}
      \setlength{\labelsep}{0.5em} } }

\newcommand{\squishend}{
    \end{list}  }

\setuldepth{cat}

\newcommand{\note}[1]{{\color{magenta}$\square$}}

\newcounter{Emp}[subsection]

\newcounter{TimeEmp}
\renewcommand{\theTimeEmp}{\arabic{TimeEmp}}
\newcommand{\TimeCNT}{\refstepcounter{TimeEmp}\textbf{\theTimeEmp}}
\newcommand{\TimeFinding}[1]{\textbf{T-{\TimeCNT\label{{#1}}}}}
\newcommand{\RefTimeFinding}[1]{\textbf{\color{blue} T-{\ref{{#1}}}}}

\newcounter{ComputeEmp}
\renewcommand{\theComputeEmp}{\arabic{ComputeEmp}}
\newcommand{\ComputeCNT}{\refstepcounter{ComputeEmp}\textbf{\theComputeEmp}}
\newcommand{\ComputeFinding}[1]{\textbf{C-{\ComputeCNT\label{{#1}}}}}
\newcommand{\RefComputeFinding}[1]{\textbf{\color{blue} C-{\ref{{#1}}}}}

\newcounter{MemoryEmp}
\renewcommand{\theMemoryEmp}{\arabic{MemoryEmp}}
\newcommand{\MemoryCNT}{\refstepcounter{MemoryEmp}\textbf{\theMemoryEmp}}
\newcommand{\MemoryFinding}[1]{\textbf{M-{\MemoryCNT\label{{#1}}}}}
\newcommand{\RefMemoryFinding}[1]{\textbf{\color{blue} M-{\ref{{#1}}}}}

\newcounter{CommunicationEmp}
\renewcommand{\theCommunicationEmp}{\arabic{CommunicationEmp}}
\newcommand{\CommunicationCNT}{\refstepcounter{CommunicationEmp}\textbf{\theCommunicationEmp}}
\newcommand{\CommunicationFinding}[1]{\textbf{CO-{\CommunicationCNT\label{{#1}}}}}
\newcommand{\RefCommunicationFinding}[1]{\textbf{\color{blue} CO-{\ref{{#1}}}}}

\newcounter{ParallelismEmp}
\renewcommand{\theParallelismEmp}{\arabic{ParallelismEmp}}
\newcommand{\ParallelismCNT}{\refstepcounter{ParallelismEmp}\textbf{\theParallelismEmp}}
\newcommand{\ParallelismFinding}[1]{\textbf{P-{\ParallelismCNT\label{{#1}}}}}

\newcounter{DataflowEmp}

\newcounter{ExecutionEmp}
\renewcommand{\theExecutionEmp}{\arabic{ExecutionEmp}}
\newcommand{\ExecutionCNT}{\refstepcounter{ExecutionEmp}\textbf{\theExecutionEmp}}
\newcommand{\ExecutionFinding}[1]{\textbf{E-{\ExecutionCNT\label{{#1}}}}}
\newcommand{\RefExecutionFinding}[1]{\textbf{\color{blue} E-{\ref{{#1}}}}}

\begin{document}
%
\title{
Survey on Characterizing and Understanding GNNs from a Computer Architecture Perspective
}
%
%
%
%

\author{Meng~Wu,
        Mingyu~Yan,~\IEEEmembership{Member,~IEEE},
        Wenming~Li, \\
        Xiaochun~Ye, 
        Dongrui~Fan,~\IEEEmembership{Senior~Member,~IEEE},
        and~Yuan~Xie,~\IEEEmembership{Fellow,~IEEE}
\IEEEcompsocitemizethanks{
\IEEEcompsocthanksitem 
Meng Wu, Mingyu Yan, Wenming Li, Xiaochun Ye, Dongrui Fan are with the Institute of Computing Technology, Chinese Academy of Sciences, Beijing, China, and also with the University of Chinese Academy of Sciences, Beijing, China
(e-mail: \{wumeng, yanmingyu, liwenming, yexiaochun, fandr\}@ict.ac.cn). 
\IEEEcompsocthanksitem 
Yuan Xie is with the Department of Electronic and Computer Engineering, Hong Kong University of Science and Technology, Hong Kong, SAR, China (e-mail: yuanxie@ust.hk).
 
\IEEEcompsocthanksitem 
Mingyu Yan is the corresponding author.

}


}

%
%

\markboth{To appear in IEEE Transactions on Parallel and Distributed Systems}
{ }
\IEEEtitleabstractindextext{%
\begin{abstract}

Characterizing and understanding graph neural networks (GNNs) is essential for identifying performance bottlenecks and facilitating their deployment in parallel and distributed systems. Despite substantial work in this area, a comprehensive survey on characterizing and understanding GNNs from a computer architecture perspective is lacking. This work presents a comprehensive survey, proposing a triple-level classification method to categorize, summarize, and compare existing efforts, particularly focusing on their implications for parallel architectures and distributed systems. We identify promising future directions for GNN characterization that align with the challenges of optimizing hardware and software in parallel and distributed systems. Our survey aims to help scholars systematically understand GNN performance bottlenecks and execution patterns from a computer architecture perspective, thereby contributing to the development of more efficient GNN implementations across diverse parallel architectures and distributed systems.

\end{abstract}

}

\maketitle

\IEEEdisplaynontitleabstractindextext

%
\IEEEpeerreviewmaketitle

\section{Introduction} \label{sec:introduction}

\IEEEPARstart{G}{raph} neural networks (GNNs) are a type of neural network specifically designed to process graph-structured data, where entities are represented as vertices and their relationships as edges. Unlike traditional deep neural networks (DNNs) that typically operate on grid-structured data such as images, GNNs leverage the connection patterns within graphs to perform tasks such as vertex classification, link prediction, and graph classification~\cite{gnn_survey,gnn_distributed_training_survey,gnn_distributed_training_survey_2}. By iteratively aggregating and transforming feature information from neighboring vertices, GNNs capture the information of both the local neighborhood and the global graph structure. This ability to naturally handle the irregular structure of graphs gives GNNs a significant advantage over traditional DNNs, offering flexibility and adaptability in tasks involving relational data~\cite{geometric_learning}. Consequently, GNNs have become powerful tools for applications including knowledge reasoning, recommendation systems, and traffic forecasting~\cite{gnn_survey,gnn_distributed_training_survey,gnn_distributed_training_survey_2,EduGraph_MOOC,GNN_Issue}.

Characterizing and understanding GNNs is crucial for identifying their performance bottlenecks and optimizing their deployment across various applications, particularly within parallel architectures and distributed systems. This process involves systematically measuring and evaluating different aspects of GNNs from a computer architecture perspective, such as compute pattern, memory pattern, and communication pattern~\cite{gcn_inf_gpu, hgnn_inf_gpu, architecture_implications_inf_CGRA}. These insights provide essential foundational knowledge and theoretical guidance for enhancing the performance and efficiency of GNN execution on parallel architectures and distributed systems. Consequently, researchers and engineers can propose effective software and hardware optimizations, such as strategies for eliminating redundant computations and exploiting parallelism, to address performance bottlenecks and facilitate the widespread adoption of GNNs. Substantial work has already been done in this area~\cite{gcn_inf_gpu, hgnn_inf_gpu, architecture_implications_inf_CGRA, GNN_Mark_gnn_training_benchmark_gpus}, highlighting the importance of such characterization in improving GNN implementations and performance within the context of parallel architectures and distributed systems.

However, there is currently a lack of comprehensive surveys specifically focused on GNN characterization. Recent surveys related to GNNs primarily concentrate on algorithmic features~\cite{gnn_survey,gnn_traffic_forecasting_survey,gnn_recommender_system_survey,DGNN_survey_2023,gnn_sampling_survey,HGNN_analysis_survey}, algorithmic and hardware acceleration techniques~\cite{GNN_acceleration_algorithmic_survey,computing_gnn_survey,FPGA_based_GCN_accelerator_survey}, and distributed training~\cite{gnn_distributed_training_survey,gnn_distributed_training_survey_2}. A consolidated survey systematically reviewing the characterization efforts for GNNs is notably absent.
Such a survey is essential for systematically organizing, summarizing, and comparing existing efforts and findings on characterizing and understanding GNNs, providing readers with a comprehensive understanding of their implications for parallel and distributed architectures. This can prevent redundant work and consolidate a vast amount of findings in one place, enabling researchers and engineers to quickly access relevant knowledge. For newcomers to the GNN research field, such a survey can serve as a comprehensive learning resource, helping them quickly grasp key findings and recent advancements.

In this work, we present a comprehensive survey on characterizing and understanding GNNs from the perspective of computer architecture, with a particular focus on their implications for parallel architectures and distributed systems. Initially, we propose a triple-level classification method that coordinates GNN model types with computer architecture perspectives. This approach first categorizes previous efforts according to the types of GNNs and then systematically summarizes the characterization findings based on their computer architecture perspectives and further subcategories. Additionally, we compare the findings across different types of GNNs to provide a holistic understanding from the perspective of computer architecture. This comparison sheds light on the different execution behaviors and characteristics of various GNN types, offering valuable insights into optimizing GNN implementations for parallel architectures and distributed systems. Finally, we identify promising directions for future research in GNN characterization, particularly in the context of parallel architectures and distributed systems.

Our contributions are summarized as follows: \par
\squishlist
    \item A triple-level classification framework for categorizing and analyzing efforts to characterize and understand GNNs from a computer architecture perspective.  
    \item A comprehensive comparison of findings across different types of GNNs, providing a holistic understanding through the lens of computer architecture.  
    \item  A discussion of promising future research directions in GNN characterization.  
\squishend

In summary, by consolidating these characterization findings, our survey aims to assist scholars in quickly, systematically, and comprehensively understanding the performance bottlenecks and diverse execution patterns of GNNs from the perspective of computer architecture. This understanding, in turn, contributes to the development of more efficient GNN implementations across a wide range of applications and parallel architectures, while also guiding hardware design for GNNs in parallel and distributed systems.
The organization of this survey is shown in Table~\ref{table:organization}.

\vspace{-10pt}

\begin{table}[!htbp]
\centering
\caption{Content guidance of this paper.} \label{table:organization}
\vspace{-10pt}
\scriptsize
\renewcommand\arraystretch{1.2} 
\setlength{\tabcolsep} {1.5mm}{
\begin{tabular}{|c|cc|}
\hline 
\textbf{Section} &  \textbf{Subsection} & \textbf{Index} \\ \hline \hline
\multirow{3}{*}{Background} 
&  Graphs and GNNs  &     \S~\ref{sec:background:graphs_gnns}\\ 
&  GNN Training and Inference  &  \S~\ref{sec:background:gnn_training_inference} \\ 
&  GNN Characterization  &  \S~\ref{sec:background:characterization} \\ \hline
\multirow{2}{*}{Taxonomy} 
&  Methodology of Taxonomy  &     \S~\ref{sec:taxonomy:taxonomy_principle}\\ 
&  Overview of Categorization  &  \S~\ref{sec:taxonomy:overview_categorization} \\ \hline
\multicolumn{2}{|c}{Characterization Efforts on SHoGNNs}  & \S~\ref{sec:characterization_efforts:SHoGNNs}\\ \hline
\multicolumn{2}{|c}{Characterization Efforts on DHoGNNs}  & \S~\ref{sec:characterization_efforts:DHoGNNs}\\ \hline
\multicolumn{2}{|c}{Characterization Efforts on SHeGNNs}  & \S~\ref{sec:characterization_efforts:SHeGNNs}\\ \hline
\multicolumn{2}{|c}{Characterization Efforts on DHeGNNs}  & \S~\ref{sec:characterization_efforts:DHeGNNs}\\ \hline
\multicolumn{2}{|c}{Summary and Comparison}  & \S~\ref{sec:summary_comparison}\\ \hline
\multicolumn{2}{|c}{Future Direction and Conclusion}  & \S~\ref{sec:direction_conclusion}\\ \hline

\end{tabular}
}
\end{table}

\vspace{-15pt}

\section{Background} \label{sec:backgound}

In this section, we introduce key concepts related to GNNs and GNN characterization.

\subsection{Graphs and GNNs} \label{sec:background:graphs_gnns}

\textit{Graphs.}
A graph is a data structure consisting of vertices connected by edges, used to represent relationships or connections between entities. Graphs can be classified based on their temporal dynamics and the diversity of their elements. 

A static graph is a type of graph where the vertices, edges, and vertex features remain unchanged throughout the computation process. In contrast, a dynamic graph changes over time, with vertices (and their features) and edges (and their features) being added, removed, or updated to reflect evolving relationships. 
Dynamic graphs come in two forms: discrete time dynamic graphs, which are series of static graph snapshots captured at specific time intervals, and continuous time dynamic graphs, which are more general and represented as timed sequences of events.
Regarding element diversity, a homogeneous graph contains vertices and edges of the same type, representing a single kind of entity and relationship, while a heterogeneous graph includes multiple types of vertices and/or edges, representing diverse entities and/or relationships. Different combinations of relationships in a heterogeneous graph form higher-order relationships, known as metapaths, which hold significant meaning. For example, the metapath author-paper-author (APA) can represent coauthorship. 

\begin{figure}[!t]
    \centering
    \includegraphics[width=1\linewidth]{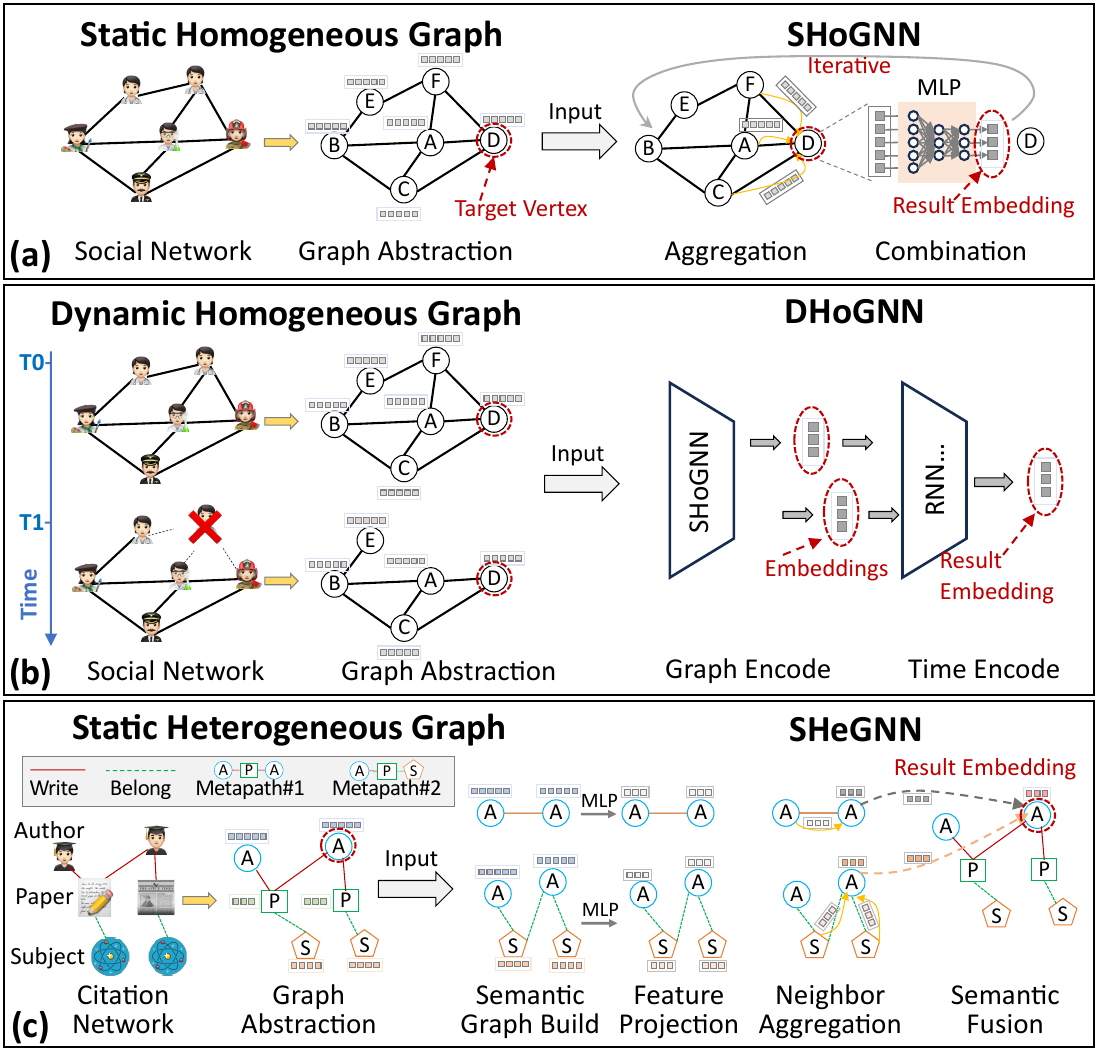}
    \vspace{-20pt}
    \caption{Graphs and GNNs: (a) Static homogeneous graph and SHoGNN, (b) dynamic homogeneous graph and DHoGNN, and (c) static heterogeneous graph and SHeGNN.}
    \label{fig:graphs_gnns}
    \vspace{-15pt}
\end{figure}

Thus, graphs can be both static or dynamic and homogeneous or heterogeneous, depending on their structural stability and the variety of their elements. For example, a static homogeneous graph, a dynamic homogeneous graph, and a static heterogeneous graph are illustrated on the left in Fig.~\ref{fig:graphs_gnns}(a), (b), and (c), respectively.

\textit{GNNs.} GNNs are designed to process graph data and can be adapted to handle different types of graphs.

Static homogeneous GNNs (SHoGNNs) are specifically designed to process static homogeneous graphs, as shown in Fig.~\ref{fig:graphs_gnns}(a). In each layer, SHoGNNs capture the relationships between vertices by aggregating the features of neighboring vertices during the aggregation stage. This aggregated features are then transformed into an embedding using a multilayer perceptron (MLP) in the combination stage. 
The aggregation stage can be further divided into three substages: scatter, apply\_edge, and gather, while the combination stage is referred to as apply\_vertex~\cite{NeuGraph} substage. 
In the scatter substage, vertex data is transmitted to its adjacent edges, constructing edge data that includes information from both the source and destination vertices. In the apply\_edge substage, a messaging function is applied to the edges using their edge data, producing intermediate tensor data associated with each edge. In the gather substage, this intermediate data is propagated along the edges and aggregated at the destination vertices.

Dynamic homogeneous GNNs (DHoGNNs) are specifically designed to handle dynamic homogeneous graphs, as shown in Fig.~\ref{fig:graphs_gnns}(b). Each layer of DHoGNNs generally consists of two stages: the graph encode stage, which uses a SHoGNN to capture structural information, and the time encode stage, which employs a recurrent neural network (RNN) to encode temporal information.

Static heterogeneous GNNs (SHeGNNs) are specifically designed to handle static heterogeneous graphs, as shown in Fig.~\ref{fig:graphs_gnns}(c). They typically involve four primary execution stages: semantic graph build, feature projection, neighbor aggregation, and semantic fusion. The semantic graph build stage partitions the heterogeneous graph into multiple semantic graphs based on a set of metapaths, with each semantic graph corresponding to a specific metapath. The feature projection and neighbor aggregation stages perform SHoGNN operations independently within each semantic graph, projecting vertex features and aggregating neighbor information. Finally, the semantic fusion stage fuses the aggregated results from all semantic graphs, integrating diverse types of high-order relationships to form comprehensive vertex embeddings.

Dynamic heterogeneous GNNs (DHeGNNs) are specifically designed to deal with dynamic heterogeneous graphs. Similar to DHoGNNs, each layer of DHeGNNs typically involves two stages: the graph encode stage, which employs a SHeGNN to capture structural information, and the time encode stage, which utilizes RNNs to encode temporal information. Due to the similarities between DHeGNNs and DHoGNNs, their illustration is omitted in Fig.~\ref{fig:graphs_gnns}.

The application of different GNN models is guided by their respective features, as described above.
SHoGNNs are particularly suitable for applications where the graph structure does not change over time, such as social network analysis or recommendation systems that operate on a fixed set of relationships. DHoGNNs, on the other hand, are more appropriate for scenarios involving temporal changes in graph data, such as traffic prediction or stock market forecasting, where both graph structure and vertex features evolve over time.
SHeGNNs can be effectively applied in domains where graphs contain diverse vertex and edge types, such as knowledge graphs or multi-modal data integration tasks. Meanwhile, DHeGNNs are suited for applications like real-time event tracking or evolving multi-relational data, where both the structure and the type of relationships between vertices vary dynamically over time.

\subsection{GNN Training and Inference} \label{sec:background:gnn_training_inference}

GNNs have two training methods including full-batch training and mini-batch training. Full-batch training is a method where the entire graph is used to perform a single update of the model parameters, providing a global perspective. It is more accurate per update but can be impractical for large graphs due to high memory and computational requirements. In contrast, mini-batch training divides the graph into smaller, manageable subgraphs (mini-batches) for each update of the model parameters. This method is more scalable and efficient for large graphs, allowing for more frequent updates and potentially faster convergence.

\begin{figure}[htbp]
    \vspace{-10pt}
    \centering
    \includegraphics[width=1\linewidth]{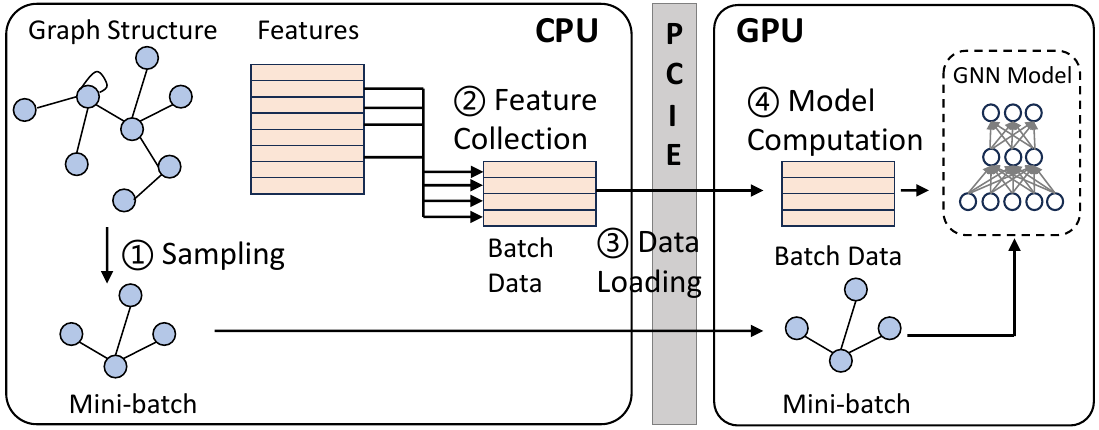}
    \vspace{-20pt}
    \caption{Workflow of mini-batch GNN training.}
    \label{fig:gnn_execution}
    \vspace{-15pt}
\end{figure}

The illustration of mini-batch training for GNNs is shown in Fig.~\ref{fig:gnn_execution}. It includes sampling, feature collection, data loading, and model computation. Sampling selects a manageable subset of vertices and edges from large graphs while considering structural, semantic, and temporal information. Feature collection gathers the features of the selected vertices and edges. Data loading organizes this information into a suitable format and loads it into GPUs, often batching for efficiency. The process of model computation involves several key steps: initializing parameters, performing forward propagation (aggregating and combining vertex features), computing the loss, calculating gradients through backward propagation, and updating the parameters. When using a multi-GPU or multi-server training platform, a gradient synchronization is needed after the backward propagation. This cycle repeats until the model converges. The GNN inference uses the trained model to perform forward propagation, making predictions on new data.

Aside from sampling mini-batches and collecting the corresponding features, the other steps of full-batch training are similar to those of mini-batch training. For simplicity, we omit the illustration figure for full-batch training of GNNs.

\subsection{GNN Characterization}\label{sec:background:characterization}

GNN characterization involves an in-depth analysis of performance aspects such as compute and memory. By using profiling tools to collect comprehensive performance metrics such as resource utilization, we can precisely identify performance bottlenecks. This detailed information is crucial for making informed decisions on optimization strategies and hardware design. For instance, profiling tools like NVIDIA Nsight Systems and Nsight Compute provide detailed insights into GPU performance. These insights can offer comprehensive guidance to accelerate GNN training and inference, enhancing efficiency and scalability.

\begin{figure}[t]
    \centering
    \includegraphics[width=0.9\linewidth]{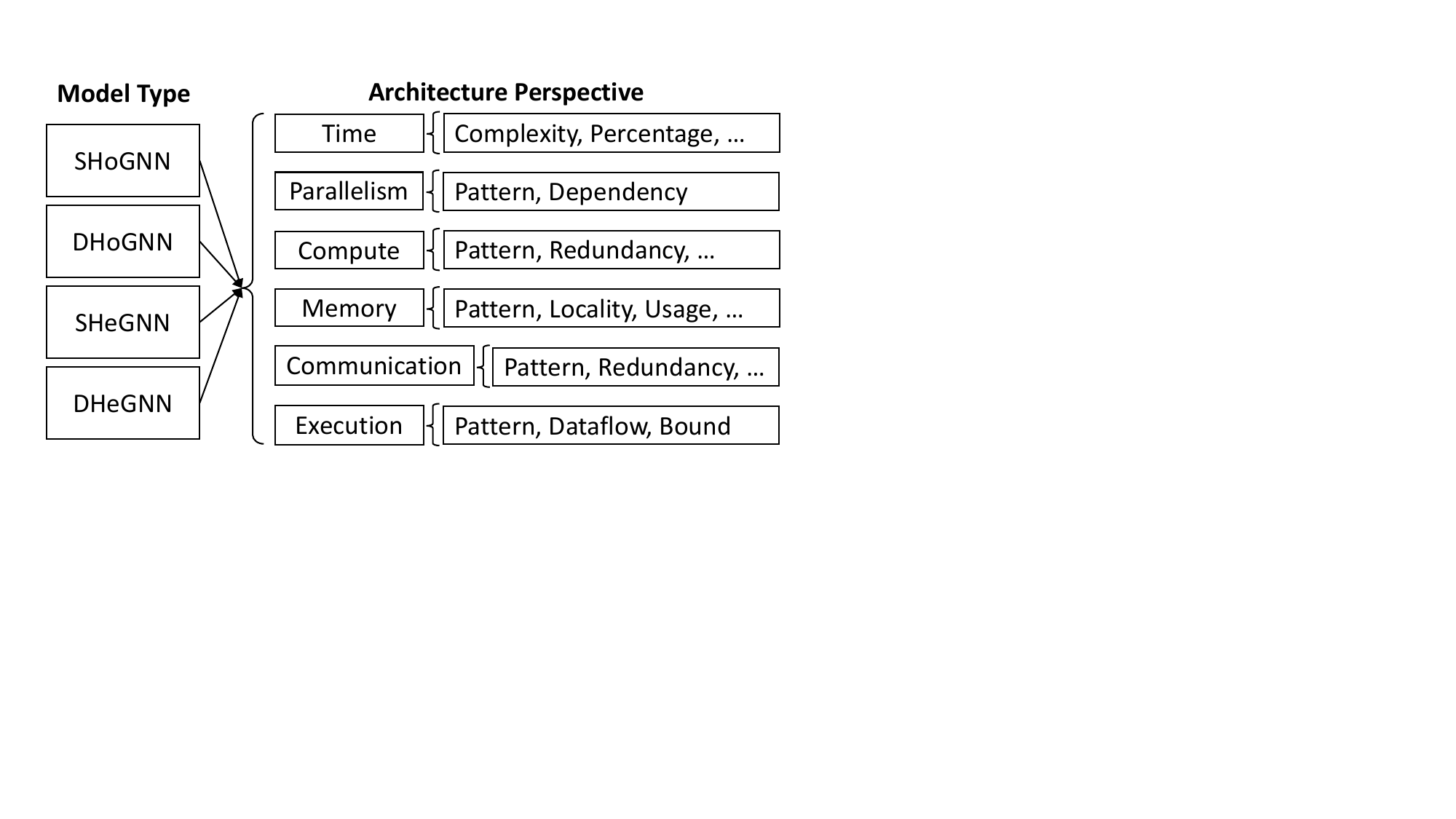}
        \vspace{-10pt}
    \caption{Methodology of taxonomy.}
    \label{fig:taxonomy}
    \vspace{-10pt}
\end{figure}

\section{Taxonomy} \label{sec:taxonomy}
In this section, we introduce the taxonomy used to classify previous characterization efforts and their findings.

\subsection{Methodology of Taxonomy} \label{sec:taxonomy:taxonomy_principle}

The methodology for classifying and analyzing prior efforts is built upon a systematic taxonomy, as shown in Fig.~\ref{fig:taxonomy}. It uses GNN model types and architecture perspectives as criteria for categorization and involves two key steps:
\squishlist
\item Step 1: Classification by GNN model type, including static homogeneous, dynamic homogeneous, static heterogeneous, and dynamic heterogeneous GNNs. This step ensures a clear distinction of efforts and findings across different GNN model types.
\item Step 2: Categorization by six architecture perspectives—Time (T), Parallelism (P), Compute (C), Memory (M), Communication (CO), and Execution (E). This step enables a deeper identification of performance bottlenecks, uncovers optimization opportunities, and distinctly highlights the focus of various findings.
\squishend

This methodology builds on prior studies that typically focus on a specific GNN model type and emphasize the significance of the following architecture perspectives in diagnosing performance challenges in parallel and distributed systems.

\textit{Time.} Understanding \textit{time complexity} provides a theoretical basis for addressing bottlenecks. Analyzing the \textit{time percentage} spent in different stages and operations reveals performance bottlenecks, making it the first step in optimization. Understanding the \textit{relationship between time and graph characteristics} enables effective preprocessing and model implementation, while the \textit{relationship between time and model setup} helps balance accuracy and speed. 

\textit{Parallelism.} Understanding the \textit{parallelism pattern} helps exploit hardware capabilities and improve scalability. Analyzing \textit{parallelism dependency} enables effective scheduling, minimizes wait times, and maximizes resource utilization.

\textit{Compute.} Understanding the \textit{compute pattern} helps optimize resource allocation and alleviate workload imbalance. Analyzing the \textit{compute pipeline} assists in identifying bottlenecks and reducing pipeline stalls. Recognizing the \textit{compute redundancy} helps reduce compute overhead. Additionally, understanding the \textit{relationship between compute amount and graph characteristics} aids in targeted optimization.

\textit{Memory.} Understanding the \textit{memory pattern} optimizes resource allocation, reduces data movement, and lowers access latency. Exploiting the \textit{data locality} enhances cache utilization and memory bandwidth use. Recognizing the \textit{memory usage} and its \textit{relationship with graph characteristics} aids in targeted optimization. Understanding the \textit{data lifetime} helps manage allocation and avoid memory fragmentation.

\textit{Communication.} Understanding the \textit{communication pattern} helps improve system performance and scalability. Analyzing the \textit{communication overhead} helps identify performance bottlenecks, while understanding the \textit{communication complexity} aids in algorithm design and resource scheduling. Recognizing the \textit{communication redundancy} helps eliminate unnecessary transfers, and understanding the \textit{communication contention} helps avoid resource contention.

\textit{Execution.} Understanding the \textit{execution pattern} aligns resource allocation with workload characteristics for higher performance. Analyzing the \textit{execution dataflow} schedules operations efficiently, reducing waiting times and computation. Understanding the \textit{execution bounds} helps pinpoint and address bottlenecks, maximizing resource utilization.

\subsection{Overview of Categorization}  \label{sec:taxonomy:overview_categorization}

Table~\ref{table:efforts_classification} presents a categorization overview of the latest efforts on GNN characterization using our proposed taxonomy. In addition, we include supplementary information in the table to offer a comprehensive review of these studies. The column labeled \textit{phase} includes training and inference phases. The column labeled \textit{software framework} lists the software frameworks used for characterization in the corresponding work. Two widely used software frameworks are PyG~\cite{PyG} and DGL~\cite{DGL}. The column of \textit{hardware platform} enumerates the hardware platforms utilized to run the software frameworks in the corresponding work. ``+'' denotes that the hardware platforms include two or more components, such as CPU + GPU. ``---'' indicates that these efforts do not provide these details. 
Guided by our taxonomy methodology, this table enables a comparative analysis of prior efforts across different GNN model types, architecture perspectives, phases, software frameworks, and hardware platforms.

\begin{table}[t]
\centering
\caption{Categorization overview of characterization efforts on GNNs.} \label{table:efforts_classification}
\vspace{-10pt}
\renewcommand\arraystretch{1.2}
\setlength\tabcolsep{2pt}%
\scriptsize
\begin{tabular}{|c|c|c|c|c|c|}
\hline
  \textbf{Work} &
  \begin{tabular}[c]{@{}c@{}}\textbf{Model} \\ \textbf{Type}\end{tabular} &
  \textbf{Perspective} &
  \textbf{Phase} &
  \begin{tabular}[c]{@{}c@{}}\textbf{Software} \\ \textbf{Framework}\end{tabular} &
  \begin{tabular}[c]{@{}c@{}}\textbf{Hardware} \\ \textbf{Platform}\end{tabular} \\ \hline \hline
  
\cite{empirical_analysis_gnn_runtime_gpus}
   & SHoGNN
   & T, M
   & Both
   & PyG
   & T4 GPU \\ \hline

 \cite{gcn_inf_gpu}
   & SHoGNN
   & T, M
   & Inference
   & PyG
   & V100 GPU  \\ \hline

\cite{distributed_gnn_training_gpu}
   & SHoGNN
   & T, M, CO
   & Training
   & PyG
   & CPU + T4 GPUs \\ \hline

\cite{GNN_Mark_gnn_training_benchmark_gpus}
   & SHoGNN
   & All
   & Training
   & GNNMark~\cite{GNN_Mark_gnn_training_benchmark_gpus}
   & CPU + V100 GPUs  \\ \hline

\cite{framework_analysis_time_memory_gnn}
   & SHoGNN
   & T, M, CO
   & Training
   & PyG, DGL
   & 2080Ti GPUs  \\ \hline

\cite{gnn_benchmarks_inf_gpus}
   & SHoGNN
   & C, M
   & Inference
   & gSuite~\cite{gnn_benchmarks_inf_gpus}
   & V100 GPU \\ \hline

\cite{parallelism_concurrency_analysis_distributed_gnn}
   & SHoGNN
   & P, C, CO, E
   & ---
   & ---
   & --- \\ \hline
                  
\cite{memory_access_patterns}
   & SHoGNN
   & M, CO
   & Training
   & PyG 
   & CPU + A100 GPU  \\ \hline

\cite{GNN_architectural_implications_GPU_TPU}
   & SHoGNN
   & T, C, M, E
   & Inference
   & PyG, DGL
   & CPU + 2080Ti GPU \\ \hline

\cite{HyGCN}
   & SHoGNN
   & T, C, M, E
   & Inference
   & PyG
   & CPU \\ \hline

\cite{nextdoor_sampling}
   & SHoGNN
   & P, T
   & Training
   & ---
   & CPU + V100 GPU \\ \hline
   
\cite{sampling_characterization}
   & SHoGNN
   & T, CO, E
   & Training
   & AliGraph~\cite{AliGraph}
   & CPU \\ \hline

\cite{huang2022characterizing}
   & SHoGNN
   & T
   & Training
   & PyG, DGL
   & CPU + 8000 GPU  \\ \hline

\cite{bgl_2023NDSI}
   & SHoGNN
   & P, T, CO
   & Training
   & DGL, Euler~\cite{Euler}
   & V100 GPUs \\ \hline

\cite{gcn_scalability_PIUMA}
   & SHoGNN
   & T
   & Inference
   & PyG
   & CPU + A100 GPU \\ \hline

\cite{pagraph_data_loading_redundancy_socc2022}
   & SHoGNN
   & T, CO
   & Training
   & DGL
   & CPU + 1080Ti GPUs \\ \hline

\cite{Cluster_GCN_KDD2019_Neibor_explosion}
   & SHoGNN
   & T, M
   & Training
   & ---
   & CPU + V100 GPU \\ \hline

\cite{GraphACT}
   & SHoGNN
   & C, M, E
   & Training
   & ---
   & CPU + P100 GPU \\ \hline

\cite{gap_optimization_complexity_gnn}
   & SHoGNN
   & C, M
   & Training
   & PyG, DGL
   & V100 GPU \\ \hline

\cite{gnn_redundancy_computation_kdd2020}
   & SHoGNN
   & C
   & Training
   & ---
   & V100 GPU \\ \hline

\cite{computational_graph_understanding}
   & SHoGNN
   & C
   & Training
   & DGL
   & 3090 GPU\\ \hline

\cite{quantization_analysis_gnn}
   & SHoGNN
   & C
   & Training
   & PyG
   & --- \\ \hline

\cite{nvmem_understanding}
   & SHoGNN
   & M
   & Training
   & PyG
   & CPU + T4 GPU \\ \hline

\cite{GNNLab}
   & SHoGNN
   & M
   & Training
   & DGL
   & CPU + V100 GPUs \\ \hline

\cite{P3_communication_characteristics}
   & SHoGNN
   & CO
   & Training
   & DGL
   & CPU + P100 GPU \\ \hline


\cite{dynamic_graph_inf_cpu_gpu}
   & DHoGNN
   & P, M, C
   & Inference
   & ---
   & CPU + A6000 GPU \\ \hline

\cite{ready_dgnn}
   & DHoGNN
   & T, M, E
   & Inference
   & PyGT~\cite{PyGT_DGNN}
   & A100 GPU \\ \hline

\cite{wang2023stag}
   & DHoGNN
   & T, C
   & Inference
   & AliGraph~\cite{AliGraph}
   & CPU + A100 GPU  \\ \hline

\cite{ipdps_dynamic_gnn_inference}
   & DHoGNN
   & P, C, M
   & Inference
   & ---
   & Titan Xp GPU \\ \hline

\cite{dgnn_booster}
   & DHoGNN
   & P, C, M
   & Inference
   & ---
   & --- \\ \hline

\cite{guan2022dynagraph}
   & DHoGNN
   & C
   & Training
   & ---
   & --- \\ \hline

\cite{wang2023tgopt}
   & DHoGNN
   & C
   & Inference
   & ---
   & V100 GPU \\ \hline

\cite{yu2023race}
   & DHoGNN
   & T, M
   & Inference
   & ---
   & --- \\ \hline

\cite{dynamic_gnn_redundancy}
   & DHoGNN
   & CO
   & Training
   & TGL~\cite{zhou2022tgl}
   & CPU + V100 GPUs \\ \hline

\cite{PiPAD_ppopp2023}
   & DHoGNN
   & P, M, CO
   & Training
   & PyGT~\cite{PyGT_DGNN}
   & CPU + V100 GPU \\ \hline

\cite{zhou2022tgl}
   & DHoGNN
   & P, E
   & Training
   & ---
   & --- \\ \hline


\cite{hgnn_inf_gpu}
   & SHeGNN
   & T, P, C, M, E
   & Inference
   & DGL
   & T4 GPU \\ \hline

\cite{HiHGNN}
   & SHeGNN
   & T, P, M, E
   & Inference
   & PyG
   & T4 GPU \\ \hline

\cite{MetaNMP}
   & SHeGNN
   & T, C, M, E
   & Inference
   & PyG
   & CPU + V100 GPU \\ \hline

\cite{HGNN_training_understanding_sc}
   & SHeGNN
   & P, C, M
   & Training
   & ---
   & --- \\ \hline

\cite{HGNN_training_kdd_DistDGLv2}
   & SHeGNN
   & M
   & Training
   & ---
   & --- \\ \hline


\cite{GraphMetaP}
   & DHeGNN
   & T, M
   & Inference
   & PyG
   & CPU + P100 GPU \\ \hline

\end{tabular}
\vspace{-20pt}
\end{table}

\section{Characterization Efforts on SHoGNNs} \label{sec:characterization_efforts:SHoGNNs}

We systematically and comprehensively summarize the findings from the characterization efforts on SHoGNNs.

\subsection{Findings on Time Perspective} \label{sec:characterization_efforts:static_gnn:time}

\begin{figure}[!t]
    \centering
    \includegraphics[width=1\linewidth]{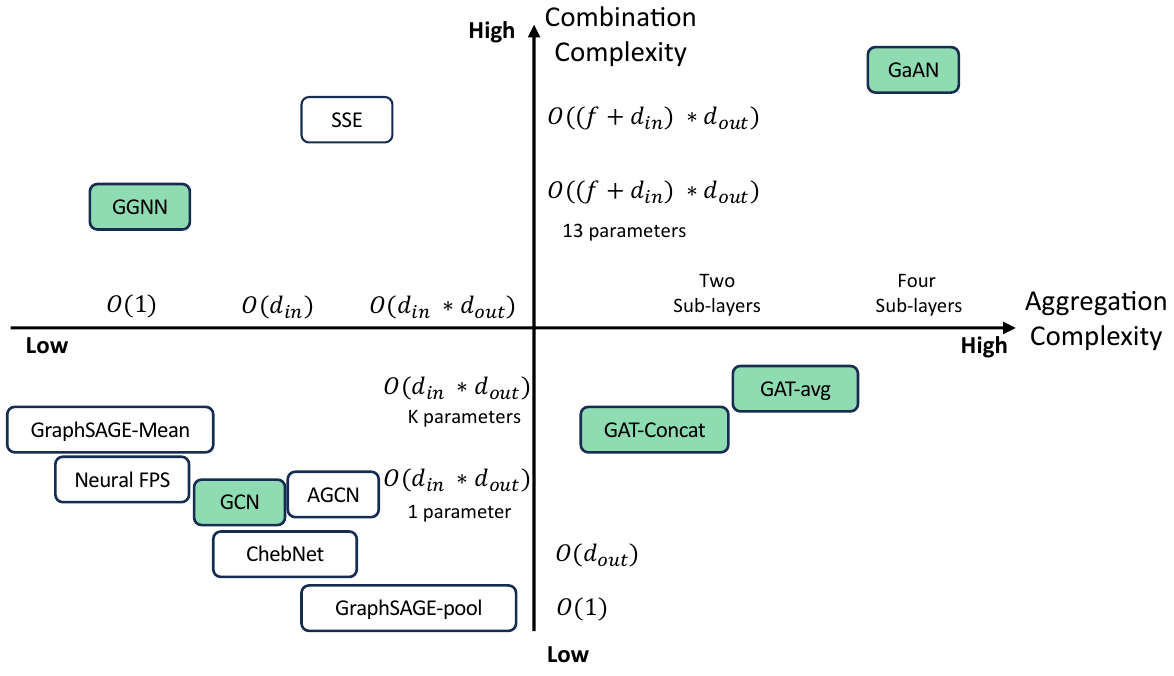}
    \vspace{-20pt}
    \caption{Time complexity quadrants of typical SHoGNNs~\cite{empirical_analysis_gnn_runtime_gpus}.}
    \label{fig:empirical_analysis_gnn_runtime_gpus:time_complexity}
    \vspace{-15pt}
\end{figure}

\textit{Time Complexity.} \TimeFinding{number:empirical_analysis_gnn_runtime_gpus:training_time_with_time_complexity} The training time is related to the model's time complexity.
\TimeFinding{number:empirical_analysis_gnn_runtime_gpus:message_time_complexity} The time complexity of the message function used in the apply\_edge substage affects the execution time of the aggregation stage~\cite{empirical_analysis_gnn_runtime_gpus, framework_analysis_time_memory_gnn}.
\TimeFinding{number:empirical_analysis_gnn_runtime_gpus:time_complexity} The time complexity for each vertex is affected by the dimensions of the input and output hidden features ($d_{in}$ and $d_{out}$) and the dimensions of the model parameters, such as the number of heads ($K$) in GAT~\cite{GAT}~\cite{empirical_analysis_gnn_runtime_gpus}. 
Accordingly, typical SHoGNNs are classified into four quadrants based on their time complexity of aggregation and combination, as shown in Fig.~\ref{fig:empirical_analysis_gnn_runtime_gpus:time_complexity}~\cite{empirical_analysis_gnn_runtime_gpus}. 
\TimeFinding{number:Cluster_GCN_KDD2019_Neibor_explosion:time} All stochastic gradient descent based training algorithms suffer from exponential time complexity with respect to the number of layers~\cite{Cluster_GCN_KDD2019_Neibor_explosion,framework_analysis_time_memory_gnn}.

\textit{Time Percentage.} \TimeFinding{number:empirical_analysis_gnn_runtime_gpus:first_layer_time} During SHoGNN training, the first layer consumes significantly more time than the second layer in model computation due to dimensionality reduction of vertex features and aggregation from more neighbors in the first layer~\cite{empirical_analysis_gnn_runtime_gpus}.
\TimeFinding{number:HyGCN:aggregation_combination_time} During SHoGNN inference, both the gather and apply\_vertex substages can dominate the execution time of each layer, varying across different datasets and models due to factors such as the variable dimensions of input and output vertex features~\cite{HyGCN}.
\TimeFinding{number:empirical_analysis_gnn_runtime_gpus:gather_apply_vertex_time} Similarly, during SHoGNN training, the aggregation stage typically dominates the execution time of each layer in model computation, except in models with high computational complexity in the apply\_vertex substage or graphs with low average vertex degrees~\cite{empirical_analysis_gnn_runtime_gpus}.
\TimeFinding{number:empirical_analysis_gnn_runtime_gpus:edge_related_calculation_time} Edge-related computations occupy most of the execution time of SHoGNN training and inference in most models~\cite{empirical_analysis_gnn_runtime_gpus}. For models with high edge computational complexity, most of time is spent on the messaging function for each edge. Otherwise, the aggregation of messages from all edges consume most of the time.
\TimeFinding{number:GNN_architectural_implications_GPU_TPU:hotspot} There is no fixed hotspot in SHoGNN training, as substage time distribution varies greatly among models~\cite{GNN_architectural_implications_GPU_TPU,GNN_Mark_gnn_training_benchmark_gpus}.

%
%

\TimeFinding{number:nextdoor_sampling:sampling_time} 
Sampling in mini-batch training and inference of SHoGNNs constitutes a significant portion of total time~\cite{nextdoor_sampling, sampling_characterization, empirical_analysis_gnn_runtime_gpus, huang2022characterizing, bgl_2023NDSI, gcn_scalability_PIUMA}. Sampling, typically performed on the CPU, can take up to 62\% of an epoch's time \cite{nextdoor_sampling}. This bottleneck worsens if the CPU is attached to multiple GPUs and cannot produce enough samples to saturate them.

%
%

\TimeFinding{number:pagraph_data_loading_redundancy_socc2022:dataloading_communication_bottleneck} Mini-batch training for SHoGNNs faces time-consuming feature collection and data loading from CPU to GPU, particularly for loading vertex features~\cite{pagraph_data_loading_redundancy_socc2022, distributed_gnn_training_gpu, gcn_scalability_PIUMA}. The study~\cite{pagraph_data_loading_redundancy_socc2022} observed that 74\% of time is spent on feature collection and data loading. This is mainly due to the large size of graph data, resulting in longer transfer time than model computation time. Moreover, using multiple GPUs within a single physical machine to speed up training proportionally increases the demand for data samples loaded from CPU to GPU.
\TimeFinding{number:pagraph_data_loading_redundancy_socc2022:collection_time} Furthermore, feature collection is CPU-intensive and takes significantly longer than CPU-GPU data movement~\cite{pagraph_data_loading_redundancy_socc2022}. When multiple GPUs are used, concurrent workers collecting features compete with samplers for CPU resources, resulting in an 88\% increase in sampling time and a 59\% increase in feature collection time compared to the single-GPU scenario~\cite{pagraph_data_loading_redundancy_socc2022}.
\TimeFinding{number:framework_analysis_time_memory_gnn:graph_classification} Data loading time for training graph classification models constitutes a significant proportion of total training time~\cite{framework_analysis_time_memory_gnn}. Unlike the general data loading focus in \RefTimeFinding{number:pagraph_data_loading_redundancy_socc2022:dataloading_communication_bottleneck}, \RefTimeFinding{number:framework_analysis_time_memory_gnn:graph_classification} specifically focuses on batching a set of small graphs into a single large batch.

\TimeFinding{number:distributed_gnn_training_gpu:gradient_synchronization} The gradient synchronization in model computation becomes more time-consuming as the number of GPUs increases~\cite{distributed_gnn_training_gpu}. Unlike DNNs, GNNs have smaller model sizes with fewer layers and shared weights across all vertices. While the gradient synchronization is expected to occupy a small part of the execution time, it takes up a larger proportion as GPU count increases due to CPU-side imbalances, causing staggered GPU computations.

\textit{Relation between Time and Graph Characteristics.} 
\TimeFinding{number:gcn_inf_gpu:linear_aggregation_combination_time} During SHoGNN inference, the execution time of the gather and apply\_vertex substages in each layer increases linearly with the dimension of the output hidden features~\cite{gcn_inf_gpu}. This is consistent with \RefTimeFinding{number:empirical_analysis_gnn_runtime_gpus:time_complexity}. \TimeFinding{number:gcn_inf_gpu:linear_combination_time} Additionally, the execution time of the apply\_vertex substage is almost proportional to the dimension of the input vertex features~\cite{gcn_inf_gpu}. This is consistent with \RefTimeFinding{number:empirical_analysis_gnn_runtime_gpus:time_complexity}.
\TimeFinding{number:empirical_analysis_gnn_runtime_gpus:linear_aggregation_combination_time} Similarly to \RefTimeFinding{number:gcn_inf_gpu:linear_aggregation_combination_time}, which focuses on SHoGNN inference, the execution time of the scatter, apply\_edge, gather, and apply\_vertex substages during SHoGNN training increases linearly with the dimension of the output hidden features~\cite{empirical_analysis_gnn_runtime_gpus}. This is consistent with \RefTimeFinding{number:empirical_analysis_gnn_runtime_gpus:time_complexity}.
\TimeFinding{number:empirical_analysis_gnn_runtime_gpus:linear_training_inference_time} The training and inference time of a layer is primarily affected by the dimensions of the input and output hidden features, increasing linearly with these dimensions~\cite{empirical_analysis_gnn_runtime_gpus}. This is consistent with \RefTimeFinding{number:empirical_analysis_gnn_runtime_gpus:time_complexity}, \RefTimeFinding{number:gcn_inf_gpu:linear_aggregation_combination_time}, \RefTimeFinding{number:gcn_inf_gpu:linear_combination_time}, and \RefTimeFinding{number:empirical_analysis_gnn_runtime_gpus:linear_aggregation_combination_time}.
\TimeFinding{number:empirical_analysis_gnn_runtime_gpus:aggregation_with_degree} The time of the aggregation stage in a layer during SHoGNN training increases linearly with the average vertex degree of the graph~\cite{empirical_analysis_gnn_runtime_gpus}.
\TimeFinding{number:empirical_analysis_gnn_runtime_gpus:portion_degree} The average vertex degree of graph affects the proportion of execution time for scatter, apply\_edge, and gather substages in the total training time~\cite{empirical_analysis_gnn_runtime_gpus}.

\textit{Relation between Time and Model Setup.} 
\TimeFinding{number:empirical_analysis_gnn_runtime_gpus:time_across_hyperparameter} Fixing other hyperparameters, each individual hyperparameter affects the training and inference time of a layer linearly~\cite{empirical_analysis_gnn_runtime_gpus}.
\TimeFinding{number:empirical_analysis_gnn_runtime_gpus:1} Sampling reduces training time per batch only when the batch size is very small. As batch size increases, the overhead of sampling and data loading causes the training time to exceed that of full-batch training~\cite{empirical_analysis_gnn_runtime_gpus}.

\subsection{Findings on Parallelism Perspective} \label{sec:characterization_efforts:static_gnn:parallelism}

\begin{figure}[!t]
    \centering
    \includegraphics[width=1\linewidth]{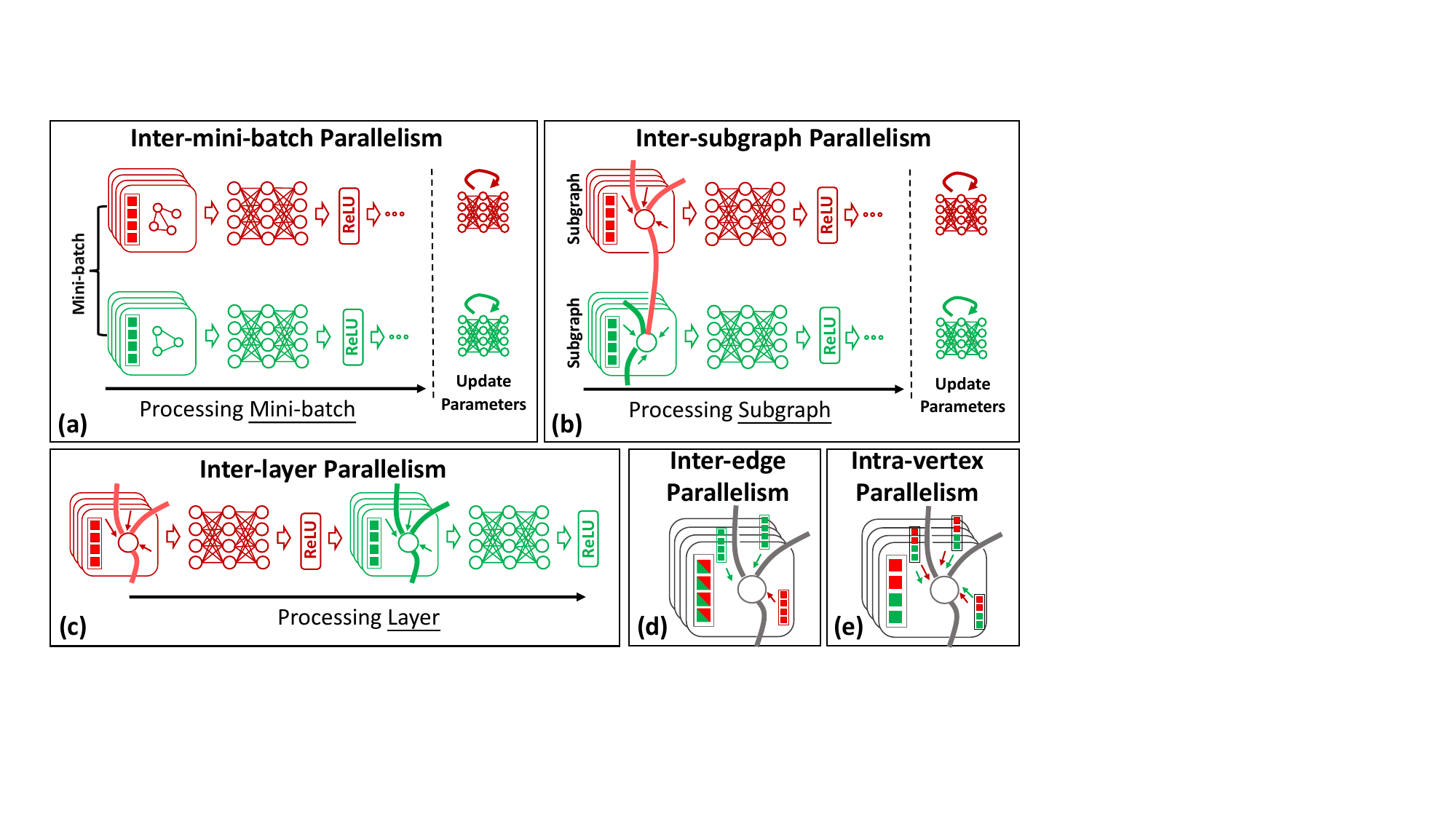}
    \vspace{-20pt}
    \caption{Parallelism examples in SHoGNNs: (a) Inter-mini-batch parallelism, (b) inter-subgraph parallelism, (c) inter-layer parallelism, (d) inter-edge parallelism, and (e) intra-vertex parallelism (different colors represent processing by different workers).}
    \label{fig:parallelism_concurrency_analysis_distributed_gnn:parallelism_taxanomy}
    \vspace{-15pt}
\end{figure}

\textit{Parallelism Pattern.} \ParallelismFinding{number:parallelism_concurrency_analysis_distributed_gnn:data_parallelism} SHoGNNs, like DNNs, exhibit data parallelism with two key variants: inter-mini-batch and inter-subgraph parallelism~\cite{parallelism_concurrency_analysis_distributed_gnn}, as shown in Fig.~\ref{fig:parallelism_concurrency_analysis_distributed_gnn:parallelism_taxanomy} (a) and (b), respectively. The former and latter respectively parallelize the processing of mini-batches and subgraphs.

\ParallelismFinding{number:parallelism_concurrency_analysis_distributed_gnn:model_parallelism} GNNs also exhibit model parallelism with several variants, such as pipeline parallelism, operator parallelism, and DNN parallelism~\cite{parallelism_concurrency_analysis_distributed_gnn}. Pipeline parallelism includes inter-layer pipeline parallelism (pipelining GNN layers as shown in Fig.~\ref{fig:parallelism_concurrency_analysis_distributed_gnn:parallelism_taxanomy} (c)) and intra-layer pipeline parallelism (pipelining the processing of a single GNN layer). Operator parallelism involves parallelizing single operators, such as inter-edge parallelism (processing different neighbors of a vertex in parallel, as shown in Fig.~\ref{fig:parallelism_concurrency_analysis_distributed_gnn:parallelism_taxanomy} (d)), inter-vertex parallelism (processing different vertices in parallel), and intra-vertex parallelism (updating different features within a single vertex in parallel, as shown in Fig.~\ref{fig:parallelism_concurrency_analysis_distributed_gnn:parallelism_taxanomy} (e))~\cite{gcn_inf_gpu}. The apply\_edge and apply\_vertex stages involve DNN operations, allowing for DNN parallelism, including DNN-pipeline parallelism and DNN-model parallelism~\cite{gcn_inf_gpu}.

\ParallelismFinding{number:nextdoor_sampling:parallelism} Sampling is ``embarrassingly parallel'' since samples are drawn independently~\cite{nextdoor_sampling}. Samples (or subgraphs) are the fundamental units of parallelism, growing in parallel by looking up the neighbors of vertices within each sample.

\textit{Parallelism Dependency.} \ParallelismFinding{number:parallelism_concurrency_analysis_distributed_gnn:parallelism}
Inter-mini-batch parallelism in SHoGNN training is more complex than in DNN training due to the numerous dependencies between data samples~\cite{parallelism_concurrency_analysis_distributed_gnn}.
\ParallelismFinding{number:inter_node:bgl_2023NDSI:resouce_contention_stage}
Resource contention exists between different training stages when they run simultaneously~\cite{bgl_2023NDSI}. Unlike DNN training, GNN training involves complex preprocessing tasks such as sampling and feature collection, which consume CPU, memory, and network bandwidth resources. Uncontrolled competition can cause poor performance, with some operations monopolizing resources and blocking others, preventing scaling with additional resources~\cite{bgl_2023NDSI}.

\ParallelismFinding{number:GNN_Mark_gnn_training_benchmark_gpus:low_ilp}
SHoGNN training exhibits low instruction-level parallelism because of numerous dependencies between instructions within a GPU warp. This is indicated by the high percentage of execution dependency stalls, averaging 29.5\%~\cite{GNN_Mark_gnn_training_benchmark_gpus}.
\ParallelismFinding{number:gcn_inf_gpu:hybrid_execution_pattern}
During the gather substage, inter-warp atomic collisions occur when multiple threads attempt to read-modify-write the same data word simultaneously\cite{gcn_inf_gpu}. Each thread processes one of the consecutive elements of a neighboring features, resulting in only inter-warp collisions~\cite{gcn_inf_gpu}.

\subsection{Findings on Compute Perspective} \label{sec:characterization_efforts:static_gnn:compute}

\textit{Compute Pattern.} \ComputeFinding{number:parallelism_concurrency_analysis_distributed_gnn:Compute} SHoGNNs exhibit a rich diversity of tensor operations, with different models utilizing a wide range of these operations~\cite{parallelism_concurrency_analysis_distributed_gnn}.
\ComputeFinding{number:GNN_architectural_implications_GPU_TPU:operation_mix} In the scatter substage, SHoGNNs can use a mix of source/destination vertex features and edge features~\cite{parallelism_concurrency_analysis_distributed_gnn}. For the gather substage, some SHoGNNs employ simple sum/mean/max operations, while others utilize more complex attention or long short-term memory operations. Most SHoGNNs employ MLP operations, though some bypass the apply\_edge or apply\_vertex substages.
\ComputeFinding{number:HyGCN:0} The gather substage involves numerous irregular computations due to the irregular connections between vertices, while the apply\_vertex substage involves regular computations, characterized by executing a matrix-vector multiply (MVM) for each vertex with a shared MLP~\cite{HyGCN}.
\ComputeFinding{number:GNN_architectural_implications_GPU_TPU:ApplyVertex_stage_compute} The apply\_edge substage can be implemented by batching all edges for parallel processing using MVM and dense–dense matrix multiply (GEMM). Similarly, the apply\_vertex substage can be implemented by batching all vertices for parallel processing using GEMM~\cite{GNN_architectural_implications_GPU_TPU}.

\ComputeFinding{number:GNNMark:integer_operation} During SHoGNN training, computation is dominated by integer operations~\cite{GNN_Mark_gnn_training_benchmark_gpus}. Results show that, on average, 64\% of the executed instructions are integer instructions, while only 28.7\% are floating-point instructions~\cite{GNN_Mark_gnn_training_benchmark_gpus}.
\ComputeFinding{number:GNN_architectural_implications_GPU_TPU:scatter_stage_compute} The scatter substage primarily involves data movement and address calculations for data accesses, preparing data for the subsequent substage~\cite{GNN_architectural_implications_GPU_TPU, gnn_benchmarks_inf_gpus}. This explains the high proportion of integer operations observed in \RefComputeFinding{number:GNNMark:integer_operation}.

\ComputeFinding{number:gap_optimization_complexity_gnn:workload_imbalance} Severe workload imbalance exists in SHoGNN training, causing long-tail effects. It arises from two primary factors: 1) The irregular connections cause significant variance in the number of neighbors per vertex~\cite{GraphACT, gap_optimization_complexity_gnn}; 2) The large dimension of vertex feature makes the computation for each neighboring vertex substantial~\cite{gcn_inf_gpu}.

\textit{Compute Pipeline.} \ComputeFinding{number:GNNMark:memory_dependency_stall} SHoGNN training is mainly stalled by memory and execution dependencies~\cite{GNN_Mark_gnn_training_benchmark_gpus}. The high percentage of memory dependency stalls, averaging 34.3\%, indicates inefficiencies in the memory subsystem serving data read requests~\cite{GNN_Mark_gnn_training_benchmark_gpus}.
\ComputeFinding{number:gcn_inf_gpu:aggregation_stall} Similarly, in SHoGNN inference, the gather substage is also stalled by memory and execution dependencies, heavily impacted by the irregular connection and load-load data dependency chains~\cite{gcn_inf_gpu}.
\ComputeFinding{number:gcn_inf_gpu:combination_stall} In contrast, the apply\_vertex substage involves intensive floating-point calculations that effectively hide memory access latency, with stalls primarily due to pipeline busy and not selected, indicating a limited number of computation units~\cite{gcn_inf_gpu}.

\textit{Compute Redundancy.} \ComputeFinding{number:gnn_redundancy_computation_kdd2020:redundancy} Significant overlap between vertex neighborhoods results in a large amount of redundant computations during the aggregation stage in SHoGNN training, due to triadic closure and clustering in real-world graphs~\cite{GraphACT, gnn_redundancy_computation_kdd2020, computational_graph_understanding, small_world_networks}. Up to 84\% of aggregations are redundant~\cite{gnn_redundancy_computation_kdd2020}.

\ComputeFinding{number:quantization_analysis_gnn:value_mean_variance_acrossing_degree} While low-precision arithmetic can reduce compute overhead~\cite{quantization_analysis_gnn}, it introduces numerical errors in SHoGNN training, particularly during the gather substage and especially at vertices with high in-degrees. The variation in output magnitudes from aggregation, which increases with in-degree, leads to greater variance in output values. Moreover, the accumulation of values at high in-degree vertices results in compounded errors that are backpropagated to many vertices, worsening gradient errors~\cite{quantization_analysis_gnn}.

\textit{Relation between Compute Amount and Graph Characteristics.}
\ComputeFinding{number:quantization_analysis_gnn:operation_acrossing_dataset} Although SHoGNNs have few parameters, the number of operations required for inference and training scales at least linearly with the graph scale and dimension of vertex feature~\cite{quantization_analysis_gnn, GNN_architectural_implications_GPU_TPU}. For instance, a 2-layer model with 32 hidden units only occupies 81KB, but processing the entire Reddit graph requires 19 billion operations~\cite{quantization_analysis_gnn, GNN_architectural_implications_GPU_TPU}.
\ComputeFinding{number:GNN_architectural_implications_GPU_TPU:2} As the graph scale increases, the total number of instructions and floating-point operations also rises~\cite{GNN_architectural_implications_GPU_TPU}. Unlike DNNs, which handle fixed-size images with a fixed amount of computation, SHoGNNs process graphs of varying scales with a variable amount of computation.

\subsection{Findings on Memory Perspective} \label{sec:characterization_efforts:static_gnn:memory}

\textit{Memory Pattern.} \MemoryFinding{number:HyGCN:indirect_access} 
The aggregation stage suffers from numerous indirect and irregular memory accesses during both SHoGNN inference~\cite{HyGCN, gcn_inf_gpu} and training~\cite{GraphACT}. This is due to the randomness of neighbor indices for each vertex, leading to extremely high numbers of L2 and L3 cache misses per kilo-instruction.
\MemoryFinding{number:gnn_benchmarks_inf_gpus:memory_dependency} During SHoGNN inference, memory dependency is the dominant stall in both the scatter and gather substages due to the irregular memory accesses~\cite{gnn_benchmarks_inf_gpus, GNN_Mark_gnn_training_benchmark_gpus}. This aligns with \RefComputeFinding{number:GNNMark:memory_dependency_stall}.

\MemoryFinding{number:HyGCN:combination} The apply\_vertex substage executes an MVM for each vertex using a shared MLP, which performs regular memory accesses, requiring only small amounts of data to be accessed from DRAM~\cite{HyGCN}. This is because MVMs are compute-intensive and the MLP weight matrix is widely shared between vertices.

\MemoryFinding{number:nvmem_understanding:sampling_memory_pattern} Sampling suffers from numerous irregular memory accesses, significantly constrained by memory latency rather than memory bandwidth~\cite{nvmem_understanding}. It predominantly involves random memory lookup operations with low computational intensity.
\MemoryFinding{number:nvmem_understanding:sampling_memory_granularity} In addition, each sampling operation involves a fine-grained eight-byte DRAM read transaction, leading to severe underutilization of DRAM read bandwidth~\cite{nvmem_understanding}.

\textit{Data Locality.} \MemoryFinding{number:GNN_Mark_gnn_training_benchmark_gpus:low_l1_cache_hitratio} SHoGNN training suffers from a low L1 D-cache hit rate across all operations~\cite{GNN_Mark_gnn_training_benchmark_gpus}. The average L1 D-cache hit rate is extremely low, at just 15\%, which is a primary cause of the stalls noted in \RefComputeFinding{number:GNNMark:memory_dependency_stall}.
\MemoryFinding{number:GNN_Mark_gnn_training_benchmark_gpus:L2_cache_hitrate} In contrast, the larger L2 cache performs significantly better, with an average hit rate of 70\%~\cite{GNN_Mark_gnn_training_benchmark_gpus}.
\MemoryFinding{number:GNN_Mark_gnn_training_benchmark_gpus:divergent_load_instructions} SHoGNN training faces a significant number of divergent load instructions across all operations~\cite{GNN_Mark_gnn_training_benchmark_gpus}. An average of 32.5\% of divergent load instructions is observed~\cite{GNN_Mark_gnn_training_benchmark_gpus}. This percentage is strongly correlated with the low L1 D-cache hit rates in \RefMemoryFinding{number:GNN_Mark_gnn_training_benchmark_gpus:low_l1_cache_hitratio}.

\MemoryFinding{number:GNNLab:inter_task_locality} Different training epochs in the same stage share most or even all of the data, resulting in excellent inter-epoch data locality in the mini-batch training~\cite{GNNLab}. For example, graph topology and features, which occupy most of the GPU memory, are shared by the sampling and feature collection stages across different epochs. \MemoryFinding{number:GNNLab:intra_task_data_locality} Within an epoch, different stages operate on different input data and share only a small amount of data within a mini-batch, leading to poor intra-epoch data locality~\cite{GNNLab}.

\MemoryFinding{number:GNN_architectural_implications_GPU_TPU:scatter_stage_memory} During SHoGNN inference, the scatter substage has a high L2 cache hit rate, despite graph processing~\cite{GraphDynS} being known for low spatial data locality~\cite{GNN_architectural_implications_GPU_TPU}. This is because SHoGNNs typically use vertex/edge features of size 32 or more, whereas graph processing like PageRank uses a size of one or two~\cite{GraphDynS}. This finding is consistent with \RefMemoryFinding{number:GNN_Mark_gnn_training_benchmark_gpus:L2_cache_hitrate} observed during SHoGNN training.

\MemoryFinding{number:gcn_inf_gpu:gather_data_locality} During SHoGNN inference, the gather substage exhibits poorer temporal data locality of neighbors’ features compared to typical graph processing~\cite{gcn_inf_gpu, gap_optimization_complexity_gnn}. This is due to the large dimension of vertex features, which increases the reuse distance of these features.
\MemoryFinding{number:gcn_inf_gpu:spatial_data_locality} However, there is high spatial data locality in the access to vertex features~\cite{gcn_inf_gpu}, which derives from accessing each large-dimension vertex feature.
\MemoryFinding{number:gcn_inf_gpu:parameters_reusability} Additionally, the model parameters in the apply\_vertex substage exhibit high reusability because SHoGNNs reuse parameters across all vertices' features~\cite{gcn_inf_gpu}.

\MemoryFinding{number:distributed_gnn_training_gpu:sampling_compete_LLC} Sampling threads compete for the last level cache (LLC) in CPU cores~\cite{distributed_gnn_training_gpu}. As more threads compete, the LLC miss rate increases, leading to longer execution time.

\textit{Memory Usage.} \MemoryFinding{number:Cluster_GCN_KDD2019_Neibor_explosion:memory} SHoGNN training exhibits exponential memory usage complexity with respect to the number of layers~\cite{Cluster_GCN_KDD2019_Neibor_explosion}.
\MemoryFinding{number:empirical_analysis_gnn_runtime_gpus:runtime_memory_usage} The peak memory usage during SHoGNN training far exceeds the size of the dataset itself~\cite{empirical_analysis_gnn_runtime_gpus}. The high memory usage for intermediate results is the main factor limiting the data scalability of the training and inference~\cite{empirical_analysis_gnn_runtime_gpus}.
\MemoryFinding{number:empirical_analysis_gnn_runtime_gpus:memory_usage_intermediate_results} Most of the memory usage comes from the intermediate results of the aggregation stage~\cite{empirical_analysis_gnn_runtime_gpus}. The forward phase generates many intermediate results cached for gradient calculation in the backward phase, increasing memory usage, explaining the high peak memory usage noted in \RefMemoryFinding{number:empirical_analysis_gnn_runtime_gpus:runtime_memory_usage}.

\MemoryFinding{number:memory_access_patterns:few_large_tensors} During SHoGNN training, a few large tensors occupy a significant portion of the allocated memory~\cite{memory_access_patterns}. The top four tensors occupy almost 80\% of the allocated memory throughout an epoch~\cite{memory_access_patterns}. This is because SHoGNNs typically have only a few layers, unlike DNNs which can have many layers, often close to 100.
\MemoryFinding{number:memory_access_patterns:largest_tensor} The largest tensor varies depending on the graph scale and dimension of the vertex feature in the input graph~\cite{memory_access_patterns}. Original and dropped-out vertex features are the largest tensors in graphs with many features, while neighboring vertex features become the largest in graphs with many edges.

\MemoryFinding{number:empirical_analysis_gnn_runtime_gpus:sampling_memory_usage} Sampling significantly reduces peak memory usage, even with large batch sizes, making it possible to train SHoGNN on large graphs using GPUs~\cite{empirical_analysis_gnn_runtime_gpus}.

\textit{Relation between Memory Usage and Graph Characteristics.} 
\MemoryFinding{number:empirical_analysis_gnn_runtime_gpus:memory_expansion_ratio_across_dimension} During SHoGNN training, the ratio of peak memory usage to memory usage after loading the dataset decreases as the dimension of input vertex features increases~\cite{empirical_analysis_gnn_runtime_gpus}. 
\MemoryFinding{number:framework_analysis_time_memory_gnn:model_complexity} The complex models, such as GAT~\cite{GAT}, require significantly more memory as batch size increases, especially on datasets with many vertices and edges due to their complex mechanisms~\cite{framework_analysis_time_memory_gnn}. 
\MemoryFinding{number:empirical_analysis_gnn_runtime_gpus:memory_usage_across_degree} During SHoGNN training and inference, peak memory usage increases linearly with the average degree of the graph~\cite{empirical_analysis_gnn_runtime_gpus}. 
\MemoryFinding{number:empirical_analysis_gnn_runtime_gpus:layer_memory_usage} Additionally, memory usage of a layer also increases linearly with the dimensions of the input and output hidden features~\cite{empirical_analysis_gnn_runtime_gpus}. 
\MemoryFinding{number:empirical_analysis_gnn_runtime_gpus:memory_usage_across_parameters} Moreover, it increases linearly with hyper-parameters~\cite{empirical_analysis_gnn_runtime_gpus}.

\textit{Data Lifetime.} \MemoryFinding{number:memory_access_patterns:lifetime} During SHoGNN training, original and dropped-out vertex features have long lifetimes, while neighboring vertex features are short-lived~\cite{memory_access_patterns}. Original vertex features are used in the dropout and even reused in the next epoch. Dropped-out vertex features are utilized in both forward and backward propagation. In contrast, neighboring vertex features are read only once.
\MemoryFinding{number:memory_access_patterns:lifetime_result} Due to the differing lifetimes of original and neighboring vertex features, the memory usage pattern varies significantly between feature-intensive and edge-intensive graphs~\cite{memory_access_patterns}.

\subsection{Findings on Communication Perspective} \label{sec:characterization_efforts:static_gnn:communication}

\textit{Communication Pattern.} \CommunicationFinding{number:sampling_characterization:sampling_communication_pattern} During mini-batch training of SHoGNN, communication in sampling exhibits a random and fine-grained pattern~\cite{sampling_characterization}. On average, 48\% of communications are for graph topology, involving indirect pointer-chasing and fine-grained (8-64 byte) transfers~\cite{sampling_characterization}. The bandwidth for these small packages is 100$\times$ lower than the peak communication bandwidth, leading to underutilized bandwidth. This observation aligns with \RefMemoryFinding{number:nvmem_understanding:sampling_memory_pattern} regarding memory access patterns.

\textit{Communication Overhead.} \CommunicationFinding{number:P3_communication_characteristics:communication} Distributed training of SHoGNN generates heavy network traffic because neighboring vertices and their features must be retrieved over the network~\cite{P3_communication_characteristics, bgl_2023NDSI}. This results in a substantial network transmission, consuming a large portion of time in communication and leading to up to 80\% GPU idle time~\cite{P3_communication_characteristics}. This observation aligns with the data loading overhead in feature collection of \RefTimeFinding{number:pagraph_data_loading_redundancy_socc2022:collection_time}. \CommunicationFinding{number:framework_analysis_time_memory_gnn:1} In addition, while training on multiple GPUs can reduce computation time, using too many GPUs may increase overall training time due to communication overhead~\cite{framework_analysis_time_memory_gnn}.

\CommunicationFinding{number:inter_node:bgl_2023NDSI:cross_partition_communication} During sampling, significant cross-partition communication overhead occurs when the graph is distributed across multiple graph store servers, necessitating frequent cross-partition communication~\cite{bgl_2023NDSI}.

\textit{Communication Complexity.} \CommunicationFinding{number:parallelism_concurrency_analysis_distributed_gnn:communication} Communication and synchronization, which rely on matrix operations that group vertex and edge features, heavily depend on the matrix representations and operations used~\cite{parallelism_concurrency_analysis_distributed_gnn}. 
\CommunicationFinding{number:pagraph_data_loading_redundancy_socc2022:linear_sampler} During mini-batch training, the demand for data samples loaded from CPU to GPU increases proportionally with the number of GPUs used within a single physical machine~\cite{pagraph_data_loading_redundancy_socc2022}.
\CommunicationFinding{number:parallelism_concurrency_analysis_distributed_gnn:scatter_communication} Communication occurs during the scatter substage if vertex features are transmitted to form edge features, resulting in $O(md)$ data transfer~\cite{parallelism_concurrency_analysis_distributed_gnn}. $m$ and $d$ are number of edges and dimension of features, respectively.
\CommunicationFinding{number:parallelism_concurrency_analysis_distributed_gnn:gather_communication} Similarly, during the gather substage, there can be $O(md)$ data movement~\cite{parallelism_concurrency_analysis_distributed_gnn}. \CommunicationFinding{number:parallelism_concurrency_analysis_distributed_gnn:other_stages_communication} 
In contrast, both apply\_edge and apply\_vertex substages do not explicitly move data~\cite{parallelism_concurrency_analysis_distributed_gnn}.

\textit{Communication Redundancy.}
\CommunicationFinding{number:pagraph_data_loading_redundancy_socc2022:1} During mini-batch training of SHoGNNs, many vertex features are loaded multiple times from the CPU to GPU because a vertex often connects to multiple training vertices, leading to repeated selection by different mini-batches~\cite{pagraph_data_loading_redundancy_socc2022}. 

\CommunicationFinding{number:GNN_Mark_gnn_training_benchmark_gpus:zero_data_transmission} During full-batch training of SHoGNNs, a significant amount of data transferred from the CPU to GPU consists of zero values~\cite{GNN_Mark_gnn_training_benchmark_gpus}. On average, 43.2\% of the data in these transfers is zero, as observed by GNNMark~\cite{GNN_Mark_gnn_training_benchmark_gpus}.

\CommunicationFinding{number:memory_access_patterns:memory_caching_strategy} Memory caching by machine learning frameworks can lead to unwanted data transfers when using CUDA Unified Memory in full-batch training~\cite{memory_access_patterns}. PyTorch and TensorFlow recycle allocated memory to avoid allocation overhead, keeping it allocated and reassigning it to new tensors. This can cause unwanted transfers from host to device memory, as the CUDA driver may map the region to host memory.
\CommunicationFinding{number:memory_access_patterns:memory_thrashing} During full-batch training, the combination of CUDA Unified Memory and large vertex feature tensors can cause thrashing during dropout, necessitating additional data transfers from host to device~\cite{memory_access_patterns}.  When memory is occupied by the large original vertex features, allocating space for the dropped-out vertex features can evict parts of the original vertex features that have not yet been used, triggering another transfer from host to device.

\textit{Communication Competition.} \CommunicationFinding{number:distributed_gnn_training_gpu:1} During mini-batch training of SHoGNNs, PCIe bandwidth competition among GPUs during data loading is significant, causing decreased bandwidth utilization~\cite{distributed_gnn_training_gpu}. As more GPUs share PCIe with the CPU, this competition intensifies, causing further declines in bandwidth utilization.

\subsection{Findings on Execution Perspective} \label{sec:characterization_efforts:static_gnn:pattern}

\textit{Execution Pattern.} \ExecutionFinding{number:parallelism_concurrency_analysis_distributed_gnn:execution} Execution of SHoGNNs can be formulated using either a graph-centric approach, which operates on individual edges or vertices, or a matrix-centric approach, which operates on matrices that group all vertex- and edge-related features~\cite{parallelism_concurrency_analysis_distributed_gnn}.
\ExecutionFinding{number:GraphACT:execution_overall} During the forward and backward of SHoGNN training formulated by the matrix-centric approach, three types of matrix multiplications $P \times Q$ are involved: (1) both $P$ and $Q$ are dense, (2) $P$ is binary sparse and $Q$ is dense, and (3) $P$ is diagonal and $Q$ is dense~\cite{GraphACT}. The type (1) operation is computationally intensive, the type (2) operation incurs many irregular memory accesses, and the type (3) operation, which scales each row of $Q$ by the diagonal elements of $P$, has a negligible contribution to the total training cost.

\ExecutionFinding{number:GNN_Mark_gnn_training_benchmark_gpus:diverse_behavior} SHoGNN training exhibits a wider range of behaviors than DNN training~\cite{GNN_Mark_gnn_training_benchmark_gpus}. Each model's characteristics can vary significantly, and even the same model may behave differently depending on the input graph and its type.
\ExecutionFinding{number:GNN_architectural_implications_GPU_TPU:common_operation_set} Despite the diversity of models, a few common operations are present in each substage, including GEMM, sparse matrix-matrix multiplication (SpMM), reductions, index selection, sorting, and element-wise operations~\cite{GNN_architectural_implications_GPU_TPU}.

\textit{Execution Dataflow.}
\ExecutionFinding{number:ieeecal2020_gcn_gpu:inter_stage_dataflow} Dataflow exists between stages for each vertex, where the output from the aggregation serves as the input for the combination~\cite{gcn_inf_gpu}.
\ExecutionFinding{number:ieeecal2020_gcn_gpu:small_dataflow} Performing the combination before the aggregation can reduce data access and computation in the aggregation, as the vertex's feature vector is typically condensed into a smaller one~\cite{gcn_inf_gpu}.

\textit{Execution Bound.} \ExecutionFinding{number:GNN_Mark_gnn_training_benchmark_gpus:1} SHoGNN training is primarily memory-bound. The average instruction per cycle is 0.55 on the GPU~\cite{GNN_Mark_gnn_training_benchmark_gpus}, indicating the memory-bound nature. 
\ExecutionFinding{number:HyGCN:aggregation_bound} 
The gather substage is memory-bound due to its extensive irregular memory accesses and low compute-to-memory access ratio~\cite{HyGCN, gcn_inf_gpu}.
\ExecutionFinding{number:HyGCN:combination_bound} 
The apply\_vertex substage is compute-bound because of its intensive computations and high degree of data reusability~\cite{HyGCN, gcn_inf_gpu}.
\ExecutionFinding{number:HyGCN:hybrid_execution_pattern} 
In summary, SHoGNNs exhibit hybrid execution bounds, combining memory-bound and compute-bound substages~\cite{HyGCN, gcn_inf_gpu}.

\ExecutionFinding{number:sampling_characterization:sampling_communication_bound} Sampling in mini-batch training is limited by communication during distributed training~\cite{sampling_characterization}. As the number of servers increases, sampling throughput scales sub-linearly due to inter-server communication overhead and the random, fine-grained communication pattern revealed by \RefCommunicationFinding{number:sampling_characterization:sampling_communication_pattern}. \ExecutionFinding{number:sampling_characterization:sampling_memory_usage} Sampling is also a memory usage bottleneck due to the traversal of the entire large-scale graph~\cite{sampling_characterization}.

\section{Characterization Efforts on DHoGNNs} \label{sec:characterization_efforts:DHoGNNs}

We systematically and comprehensively summarize the findings from the characterization efforts on DHoGNNs.

\subsection{Findings on Time Perspective} \label{sec:characterization_efforts:dynamic_gnn:time}

\textit{Time Percentage.} \TimeFinding{number:ready_dgnn:time_percentage} During DHoGNN inference, graph encode and time encode stages consistently remain the most time-consuming, though execution time varies significantly depending on the dataset and model combinations~\cite{ready_dgnn}. For example, the graph encode stage can take up as much as 80.4\% of the total time in some scenarios, while in others, it may constitute only 39.6\%, as noted in work~\cite{ready_dgnn}.

\textit{Relation between Time and Graph Characteristics.} \TimeFinding{number:ready_dgnn:time_relation} During DHoGNN inference, processing all modifications when the graph changes incurs substantial additional execution time~\cite{ready_dgnn}.
\TimeFinding{number:wang2023stag:time} Additionally, updating vertex embeddings is particularly time-consuming in high-connectivity graphs and deep models due to the neighbor explosion~\cite{wang2023stag}. When a vertex is updated, the number of affected vertices grows exponentially with the number of layers and the graph's connectivity. For instance, in the Reddit dataset using a three-layer SHoGNN, it can take over 10 seconds to recalculate vertex embeddings following minor graph changes~\cite{wang2023stag}.

\subsection{Findings on Parallelism Perspective} \label{sec:characterization_efforts:dynamic_gnn:parallelism}

\textit{Parallelism Pattern.} \ParallelismFinding{number:PiPAD_ppopp2023:parallelism_cross_snapshot} The separation of time-independent and time-dependent encode in DHoGNNs enables the parallel execution of graph encode for different snapshots~\cite{PiPAD_ppopp2023}.

\textit{Parallelism Dependency.} \ParallelismFinding{number:zhou2022tgl:parallelism} Training DHoGNNs with large batch sizes is challenged by temporal dependencies due to different edges connected to the same vertices~\cite{zhou2022tgl}.

\ParallelismFinding{number:dynamic_graph_inf_cpu_gpu:temporal_data_dependencies} DHoGNN inference faces temporal data dependencies between different snapshots~\cite{PiPAD_ppopp2023}, vertex embeddings, and edge embeddings~\cite{dynamic_graph_inf_cpu_gpu, dgnn_booster}. For discrete-time dynamic graphs, all vertex and edge embeddings for a snapshot are processed simultaneously, with the next snapshot processed only after completing all updates. For continuous-time dynamic graphs, vertex and edge embeddings are updated serially according to event times.

\ParallelismFinding{number:ipdps_dynamic_gnn_inference:stage_dependency} During DHoGNN inference, asynchronous graph change events at varying rates create intrinsic sequential dependencies in temporal neighbor sampling and vertex updates, necessitating the processing of small batches~\cite{ipdps_dynamic_gnn_inference}.

\subsection{Findings on Compute Perspective} \label{sec:characterization_efforts:dynamic_gnn:compute}

\textit{Compute Pattern.} \ComputeFinding{number:ipdps_dynamic_gnn_inference:Compute} During DHoGNN inference, the computation is primarily dominated by the graph encode stage, which aggregates data from selected temporal neighbors~\cite{ipdps_dynamic_gnn_inference}.
\ComputeFinding{number:dgnn_booster:compute} Additionally, intensive matrix multiplies and complex mathematical operations are involved~\cite{dgnn_booster}.

\textit{Compute Redundancy.} \ComputeFinding{number:guan2022dynagraph:compute} Similar to SHoGNN training, redundant neighbor aggregation occurs in DHoGNN training. This redundancy arises because DHoGNNs execute repeated aggregation before any linear transformation~\cite{guan2022dynagraph}.

\ComputeFinding{number:wang2023tgopt:edge_redundant_compute} During DHoGNN inference, handling source and destination vertices within a batch leads to redundant vertex embedding computations~\cite{wang2023tgopt}. First, duplicates may exist in the initial batch that a model receives as input at the starting layer. Second, as the model recursively computes embeddings, pooling neighbors of all target vertices in the batch can lead to duplicates. Lastly, at layer 0, where vertex features are retrieved, static vertex features make target timestamps irrelevant, so only the target vertex is checked, potentially introducing duplicates that weren't there before. These redundancies can account for up to 55\% of batch computations and increase through the layers~\cite{wang2023tgopt}.

\ComputeFinding{number:wang2023tgopt:phase_redundant_compute} During DHoGNN inference, calculating embeddings for a given vertex and timestamp often involves revisiting the same temporal neighborhoods as in prior time steps, leading to redundant recalculations~\cite{wang2023stag,wang2023tgopt}. In some dynamic graphs, up to 89.9\% of the total embeddings generated over time are repeated computations~\cite{wang2023tgopt}.

\ComputeFinding{number:wang2023tgopt:time_delta_compute} During DHoGNN inference, the time-encoding operation is often invoked with the same time delta values, resulting in repeated computations~\cite{wang2023tgopt}. Since the parameters for the time encoder are fixed during inference, these values can be precomputed once and reused. Additionally, time deltas tend to follow a power-law distribution and cluster near zero due to the use of most-recent sampling.

\subsection{Findings on Memory Perspective} \label{sec:characterization_efforts:dynamic_gnn:memory}

\textit{Memory Pattern.} \MemoryFinding{number:yu2023race:memory} Similar to SHoGNN inference, DHoGNN inference experiences serious irregular memory access due to the time-evolving graph topology and irregular connections between vertices~\cite{yu2023race, dgnn_booster, dynamic_graph_inf_cpu_gpu, ready_dgnn}.  
\MemoryFinding{number:ready_dgnn:memory_pattern} Additionally, memory accesses to source vertices and their neighbors are not only irregular but also dynamic, changing with the graph topology across snapshots~\cite{ready_dgnn}.

\MemoryFinding{number:ready_dgnn:data_movement} During DHoGNN inference, the graph encode and time encode stages are highly data-dependent and often executed alternately, leading to significant data movement for intermediate results~\cite{ready_dgnn}.
\MemoryFinding{number:dynamic_graph_inf_cpu_gpu:sampling_memory_pattern} During DHoGNN inference, sampling suffers from irregular memory accesses due to the consideration of both temporal and structural information~\cite{ipdps_dynamic_gnn_inference}. Thus, it often needs sorting operations.

\textit{Data Locality.} \MemoryFinding{number:PiPAD_ppopp2023:data_locality} During DHoGNN training, the overlap among snapshots in adjacent sliding windows provides opportunities to reuse neighbor aggregation results~\cite{PiPAD_ppopp2023}. The sliding window mechanism feeds multiple continuous snapshots to DHoGNN with a forward stride size. The topology change rate among adjacent snapshots in real-life dynamic graphs is generally limited (around 10\% on average), resulting in significant snapshot overlaps~\cite{PiPAD_ppopp2023}.

\MemoryFinding{number:yu2023race:data_locality} DHoGNN inference exhibits strong spatial data locality, with most accesses to vertex features concentrated on a small set of vertices within each snapshot or across successive snapshots~\cite{yu2023race}. More than 73.4\% of accesses refer to the features of just 0.5\% of vertices, and over 80.3\% of these accesses remain consistent across successive snapshots~\cite{yu2023race}. This occurs because real-world dynamic graphs show strong temporal similarity, with 86.7\%–95.9\% of vertices remaining unaffected between consecutive snapshots, as each batch of graph updates impacts only a small set of vertices~\cite{yu2023race}.
 
\MemoryFinding{number:ready_dgnn:poor_data_locality} The time encode stage in DHoGNN inference involves substantial data dependencies, as processing each vertex requires features from a previous snapshot, resulting in poor data locality~\cite{ready_dgnn}. 
\MemoryFinding{number:ready_dgnn:memory} However, the parameters of the time-encoding model are fully shared by all vertices in each snapshot, showing unique inter-vertex locality~\cite{ready_dgnn}.

\textit{Memory Usage.} \MemoryFinding{number:dgnn_booster:memory} During DHoGNN inference, the time-evolving vertex embeddings cause significant memory consumption~\cite{dgnn_booster}.
\MemoryFinding{number:dynamic_graph_inf_cpu_gpu:sampling_memory_usage} Additionally, sampling additional neighboring vertices increases memory usage~\cite{dynamic_graph_inf_cpu_gpu}.

\subsection{Findings on Communication Perspective} \label{sec:characterization_efforts:dynamic_gnn:communication}

\textit{Communication Pattern.} \CommunicationFinding{number:dynamic_gnn_redundancy:time} In a large-scale GPU cluster, DHoGNN training requires collective operations, such as all-to-all or all-gather, to manage dependencies and aggregate updates~\cite{dynamic_gnn_redundancy}. This is necessary because temporal dependencies and parameters are distributed across GPUs.

\textit{Communication Overhead.} \CommunicationFinding{number:PiPAD_ppopp2023:communication} During DHoGNN training, continuous updates of graph snapshots cause significant communication overheads~\cite{PiPAD_ppopp2023}. Communications account for 39\% of the total time~\cite{PiPAD_ppopp2023}.
\CommunicationFinding{number:dynamic_graph_inf_cpu_gpu:3} Similarly, during DHoGNN inference, frequent data exchanges between the CPU and GPU, caused by the continuous transfer of evolving graph topology and embeddings, lead to substantial data movement~\cite{dynamic_graph_inf_cpu_gpu}. 
In contrast, SHoGNNs transfer graph topology and embeddings to the GPU only once~\cite{dynamic_graph_inf_cpu_gpu}. 
\CommunicationFinding{number:dynamic_gnn_redundancy:time_overhead} 
Thus, communication time can be substantial, reaching up to 2.7$\times$ that of computation within a mini-batch~\cite{dynamic_gnn_redundancy}.

\textit{Communication Redundancy.}
\CommunicationFinding{number:dynamic_gnn_redundancy:communication} DHoGNN training relies on the latest temporal dependencies of all vertex interactions, causing significant redundancy in cross-GPU transfers for vertices' embeddings~\cite{dynamic_gnn_redundancy}. DHoGNNs dispatch current state dependencies for computation and then aggregate the latest dependencies, resulting in many redundant communications. Over 80\% of vertices' embeddings are repeatedly transferred due to multiple interconnections~\cite{dynamic_gnn_redundancy}.

\subsection{Findings on Execution Perspective} \label{sec:characterization_efforts:dynamic_gnn:execution}

\textit{Execution Pattern.} \ExecutionFinding{number:zhou2022tgl:execution} During DHoGNN training, sampling neighbors on dynamic graphs is complex as it must consider the timestamps of the neighbors~\cite{zhou2022tgl}.
\ExecutionFinding{number:ready_dgnn:execution} During DHoGNN inference, the two encode stages share operational similarities, both performing a series of MVMs~\cite{ready_dgnn}.

\section{Characterization Efforts on SHeGNNs} \label{sec:characterization_efforts:SHeGNNs}

We systematically and comprehensively summarize the findings from the characterization efforts on SHeGNNs.

\subsection{Findings on Time Perspective} \label{sec:characterization_efforts:heterogeneous_gnn:time}

\textit{Time Percentage.} \TimeFinding{number:hgnn_inf_gpu:hgnn_four_stages_execution_time:reduction_tree} During SHeGNN inference, operations with reduction-tree-based computational graphs dominate the execution time, as each resulting element is computed using this type of graph~\cite{hgnn_inf_gpu}. \TimeFinding{number:hgnn_inf_gpu:hgnn_four_stages_execution_time:kernel_type_time} Additionally, four major operation types consume most of the execution time across the feature projection, neighbor aggregation, and semantic fusion stages. These are dense-dense matrix multiplication, graph-topology-based operations, element-wise operations, and data rearrangement~\cite{hgnn_inf_gpu}.

\TimeFinding{number:MetaNMP:sgb_time} Metapath instance matching for semantic graph building is extremely time-consuming~\cite{MetaNMP}. It takes 8129$\times$ longer than the total inference time, which can be amortized during training but is intolerable for real-time inference.
\TimeFinding{number:hgnn_inf_gpu:fp_stage_time} During SHeGNN inference, the feature projection stage's execution time is primarily dominated by dense-dense matrix multiplication operations~\cite{hgnn_inf_gpu}.
\TimeFinding{number:hgnn_inf_gpu:NA_domination_time} During SHeGNN inference, the neighbor aggregation stage consumes the majority of execution time due to the time-consuming process of aggregating neighboring feature vectors for each vertex~\cite{hgnn_inf_gpu, MetaNMP, HiHGNN}.
\TimeFinding{number:hgnn_inf_gpu:hgnn_four_stages_execution_time:3} During SHeGNN inference, the execution time in the neighbor aggregation stage is largely consumed by graph-topology-based and element-wise operations~\cite{hgnn_inf_gpu}.
\TimeFinding{number:hgnn_inf_gpu:hgnn_four_stages_execution_time:4} During SHeGNN inference, the semantic fusion stage's execution time is primarily consumed by dense-dense matrix multiplication, element-wise operations, and data rearrangement operations~\cite{hgnn_inf_gpu}.

\textit{Relation between Time and Graph Characteristics.} \TimeFinding{number:hgnn_inf_gpu:hgnn_four_stages_execution_time:increasing_neighbor_aggregation_time} During SHeGNN inference, the execution time of the neighbor aggregation stage increases not only with the average number of neighbors, as in SHoGNNs, but also with the number of metapaths~\cite{hgnn_inf_gpu}. Each additional metapath introduces an extra semantic graph, requiring more time for neighbor aggregation.
\TimeFinding{number:hgnn_inf_gpu:hgnn_four_stages_execution_time:increasing_time} Similarly, the total execution time increases with the number of metapaths~\cite{hgnn_inf_gpu}. More metapaths result in additional semantic graphs, needing extra time for both neighbor aggregation and semantic fusion.

\subsection{Findings on Parallelism Perspective} \label{sec:characterization_efforts:heterogeneous_gnn:parallelism}

\textit{Parallelism Pattern.} \ParallelismFinding{number:hgnn_inf_gpu:parallelism} In addition to inter-vertex, intra-vertex, and inter-edge parallelism~\cite{gcn_inf_gpu,HiHGNN}, inter-semantic-graph parallelism exists in the neighbor aggregation stage~\cite{hgnn_inf_gpu,HGNN_training_understanding_sc}. This parallelism arises from the independent neighbor aggregation for each semantic graph. As the number of metapaths increases, more opportunities for parallel execution are offered, as indicated by \RefComputeFinding{number:HGNN_training_understanding_sc:compute_amount}~\cite{HGNN_training_understanding_sc}.

\subsection{Findings on Compute Perspective} \label{sec:characterization_efforts:heterogeneous_gnn:compute}

\textit{Compute Redundancy.} \ComputeFinding{number:MetaNMP:redundant_computations} During SHeGNN inference, aggregating vertex features in the neighbor aggregation stage along metapath instances results in substantial redundant computations~\cite{MetaNMP}. The features of all vertices within each metapath instance are aggregated, leading to repeated aggregations for vertices that frequently appear in multiple metapath instances.

\textit{Relation between Compute Amount and Graph Characteristics.} \ComputeFinding{number:HGNN_training_understanding_sc:compute_amount} During SHeGNN training, as the number of metapaths increases, the compute amount grows linearly~\cite{HGNN_training_understanding_sc}. More metapaths cause additional semantic graphs, requiring extra computations. This aligns with \RefTimeFinding{number:hgnn_inf_gpu:hgnn_four_stages_execution_time:increasing_neighbor_aggregation_time} and \RefTimeFinding{number:hgnn_inf_gpu:hgnn_four_stages_execution_time:increasing_time}.

\subsection{Findings on Memory Perspective} \label{sec:characterization_efforts:heterogeneous_gnn:memory}

\textit{Memory Pattern.} 
\MemoryFinding{number:MetaNMP:match_irregular_access} The metapath instance matching for semantic graph building involves a large number of irregular memory accesses, including irregular edge data accesses to match metapaths~\cite{MetaNMP}.
\MemoryFinding{number:hgnn_inf_gpu:neighbor_aggregation_irregular_access} During SHeGNN inference, memory accesses to neighboring features in the neighbor aggregation stage are highly dependent on the graph's irregular connections, leading to irregular memory access patterns~\cite{hgnn_inf_gpu, MetaNMP}.

\textit{Data Locality.}
\MemoryFinding{number:hgnn_inf_gpu:locality} During SHeGNN inference, as the length of the metapath increases, the adjacency matrix for each semantic graph becomes less sparse, indicating improved locality of vertex features~\cite{hgnn_inf_gpu}. Longer metapaths lead to an exponential increase in the number of metapath instances~\cite{MetaNMP}, allowing for more neighbors to be found for each vertex.
\MemoryFinding{number:HiHGNN:1} During SHeGNN inference, projected vertex features can be reused across semantic graphs in the neighbor aggregation~\cite{HiHGNN}. Each type of vertex is projected only once, enabling the reuse of projected features if semantic graphs have significant overlap in vertex types.

\textit{Memory Usage.}
\MemoryFinding{number:HGNN_training_understanding_sc:memory} During SHeGNN training, memory usage of intermediate results increases with the number of metapaths~\cite{HGNN_training_understanding_sc}. With potentially hundreds of metapaths, this can lead to fragmented memory allocation and many small tensors due to imbalances in the scales of semantic graphs, resulting in excessive memory consumption.

\MemoryFinding{number:MetaNMP:memory_usage} During SHeGNN inference, storing all metapath instances for use in the neighbor aggregation and semantic fusion causes substantial memory usage~\cite{MetaNMP}. It requires on average 240$\times$ more memory than the graph data itself.

\subsection{Findings on Communication Perspective} \label{sec:characterization_efforts:heterogeneous_gnn:communication}

\textit{Communication Pattern.} 
\MemoryFinding{number:HGNN_training_kdd_DistDGLv2:remote_data_access} During SHeGNN training, communications are imbalanced due to variations in the number of vertices and edges in mini-batches, as well as the different dimensions of features for each vertex type~\cite{HGNN_training_kdd_DistDGLv2}.

\subsection{Findings on Execution Perspective} \label{sec:characterization_efforts:heterogeneous_gnn:execution}

\begin{figure}[!t]
 	  \centering
    	\includegraphics[width = 0.9\linewidth]{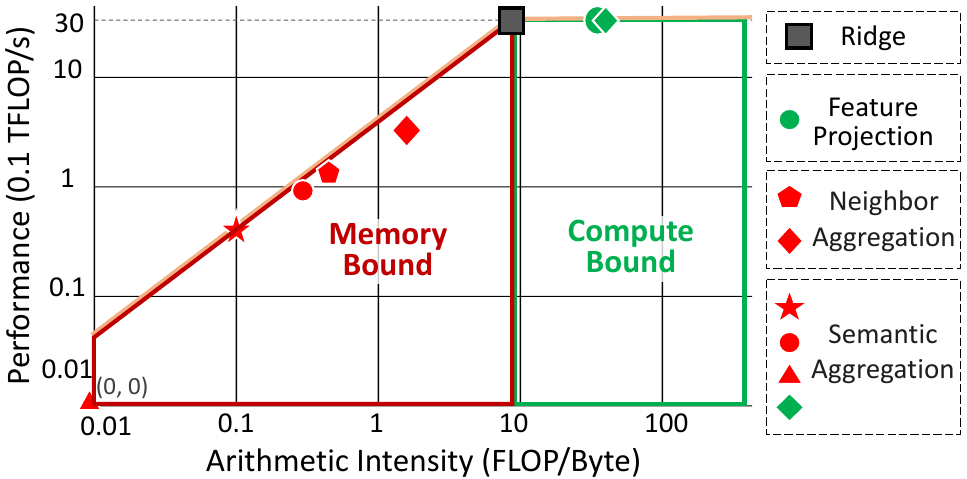}
    	\vspace{-10pt}
    	\caption{Kernels in single-precision floating point Roofline on T4 GPU (different shapes represent kernels in the corresponding stages)~\cite{hgnn_inf_gpu}.}
    	\label{fig:ieeecal2022_hgnn:rootline}
     \vspace{-10pt}
\end{figure}

\textit{Execution Bound.} \ExecutionFinding{number:MetaNMP:sgb_bound} During SHeGNN inference, the metapath instance matching for semantic graph building on the CPU is memory-bound due to irregular memory accesses and a low compute-to-memory access ratio~\cite{MetaNMP}.
\ExecutionFinding{number:hgnn_inf_gpu:FP_bound} Additionally, the feature projection stage is primarily constrained by compute resources~\cite{hgnn_inf_gpu,MetaNMP,HiHGNN}. Dense-dense matrix
multiply dominates its execution time~\cite{hgnn_inf_gpu}. As illustrated in Fig.~\ref{fig:ieeecal2022_hgnn:rootline}, the kernel in this stage is compute-bound according to the Rootline model~\cite{hgnn_inf_gpu}.
\ExecutionFinding{number:hgnn_inf_gpu:NA_bound} In contrast, the neighbor aggregation stage is primarily memory-bound~\cite{hgnn_inf_gpu,MetaNMP,HiHGNN}. This stage involves graph-topology-based and element-wise operations, which consume most of the execution time and result in irregular memory accesses and a low compute-to-memory access ratio~\cite{hgnn_inf_gpu,MetaNMP,HiHGNN}. As shown in Fig.~\ref{fig:ieeecal2022_hgnn:rootline}, the kernels in this stage are memory-bound according to the Rootline model~\cite{hgnn_inf_gpu}.
\ExecutionFinding{number:hgnn_inf_gpu:SA_bound} Furthermore, the semantic fusion stage is initially compute-bound but transitions to being memory-bound~\cite{hgnn_inf_gpu,MetaNMP,HiHGNN}. This stage first computes attention weights through dense-dense matrix multiplication and then aggregates feature vectors using element-wise operations with the attention weights~\cite{hgnn_inf_gpu}. As depicted in Fig.~\ref{fig:ieeecal2022_hgnn:rootline}, the kernels in this stage are compute-bound or memory-bound according to the Rootline model~\cite{hgnn_inf_gpu}.

\section{Characterization Efforts on DHeGNNs} \label{sec:characterization_efforts:DHeGNNs}

We systematically and comprehensively summarize the findings from the characterization efforts on DHeGNNs. Due to the limited amount of existing work on DHeGNNs, the summary and discussion of related findings are focused on two key aspects.

\subsection{Findings on Time Perspective} \label{sec:characterization_efforts:dhe_gnn:time}

\textit{Time Percentage.} \TimeFinding{number:GraphMetaP:1} As with SHeGNN inference, metapath instance matching for semantic graph building in DHeGNN is extremely time-consuming, leading to intolerable overhead for real-time inference~\cite{GraphMetaP}. It involves vertex-order matching of metapaths, which is highly time-intensive. When graph is updated, the instances must be rematched.

\subsection{Findings on Memory Perspective} \label{sec:characterization_efforts:dhe_gnn:memory}

\textit{Data Locality.} \MemoryFinding{number:GraphMetaP:data_locality} Most changes to metapath instances are concentrated within a small portion of the graph~\cite{GraphMetaP}, revealing high-degree reusability of metapath instances during semantic graph building. Since the majority of graph updates affect only a small subset of vertices and edges, only 10\% to 20\% of metapath instances are impacted~\cite{GraphMetaP}.

\textit{Memory Usage.} \MemoryFinding{number:GraphMetaP:memory_usage} Metapath instance rematching required after graph changes exacerbates the already substantial storage demands for millions of instances~\cite{GraphMetaP}.

\section{Summary and Comparison} \label{sec:summary_comparison}

We systematically summarize and compare the findings across different GNN types from six key perspectives.

\textit{Time.} 
In terms of dominant stages, SHoGNNs are primarily time-consuming in data loading, sampling, and aggregation stages. DHoGNNs mainly consume time in graph encode and time encode stages. SHeGNNs and DHeGNNs both find semantic graph build and neighbor aggregation stages to be the most time-consuming.
Regarding impact factors on execution time, SHoGNNs' execution time increases with feature dimension, vertex degree, and hyperparameters. For DHoGNNs, execution time further increases with the number of updated vertices and edges, compared to SHoGNNs. Similarly, SHeGNNs' execution time increases with the number of metapaths, while DHeGNNs' execution time also rises with the number of updated vertices and edges, compared to SHoGNNs.

\textit{Parallelism.} 
In terms of parallelism pattern, all four types of GNNs can possess parallelism in mini-batches, subgraphs, layers, inter-vertex processing, inter-edge processing, and intra-feature processing. Additionally, SHeGNNs and DHeGNNs can leverage parallelism across different semantic graphs, while DHoGNNs and DHeGNNs can exploit parallelism across snapshots.
Regarding parallelism dependency, all four types of GNNs can encounter dependencies when updating the same vertex in parallel and in the sequential execution of each stage. Moreover, DHoGNNs and DHeGNNs face additional challenges with inter-event parallelism dependency and inter-snapshot parallelism dependency in the time encode stage.

\textit{Compute.} 
In terms of compute pattern, all four types of GNNs rely on graph-topology-based and MLP-based operations to capture structural and semantic information. Additionally, DHoGNNs and DHeGNNs require RNN-based operations for time encoding.
Regarding compute redundancy, SHoGNNs suffer from redundant aggregation operations due to overlapping neighborhoods among target vertices. DHoGNNs further face redundant graph encoding operations on unchanged vertices and the same temporal neighborhoods, compared to SHoGNNs. SHeGNNs experience additional redundancy in neighbor aggregation due to overlapping vertices between metapath instances. DHeGNNs further suffer from redundant graph encoding operations on unchanged vertices and the same temporal neighborhoods, compared to SHeGNNs.
In terms of impact factors on compute amount, all four types of GNNs see an increase in the number computations with the growing scale of graphs and the dimension of features. The number of computations for SHeGNNs and DHeGNNs further increases with the number of metapaths.

\textit{Memory.} 
In terms of memory pattern, all four types of GNNs face irregular memory accesses due to graph-topology-based or RNN-based operations and regular memory accesses due to MLP-based operations. Additionally, DHoGNNs and DHeGNNs deal with dynamic memory accesses as graph-topology-based operations are performed on evolving graphs.
Regarding data locality, all four types of GNNs exhibit data locality of input vertex features across epochs, data locality within vertex features, and reusability of parameters. DHoGNNs and DHeGNNs further exhibit spatial data locality within each snapshot or across successive snapshots. SHeGNNs and DHeGNNs show reusability of projected features across semantic graphs.
In terms of memory usage, SHoGNNs primarily use memory for original and dropped-out vertex features, while DHoGNNs primarily use memory for time-evolving vertex features. SHeGNNs and DHeGNNs primarily use memory for metapath instances.
Regarding the impact factors on memory usage, memory usage for all four types of GNNs increases with the dimension of features, vertex degree, and hyperparameters. Memory usage in SHeGNNs and DHeGNNs also increases with the number of metapaths.

\textit{Communication.} 
In terms of communication pattern, all four types of GNNs face random transfers due to irregular connections and potential imbalances due to the non-uniform distribution of vertices.
Regarding primary communication data, SHoGNNs and SHeGNNs primarily transfer features of neighboring vertices and mini-batches. In contrast, DHoGNNs and DHeGNNs mainly transfer evolving graph topology and vertex features.
In terms of redundancy, all four types of GNNs can introduce redundant transfers of vertex features. DHoGNNs and DHeGNNs further introduce the latest features of vertices.

\textit{Execution.} 
In terms of execution pattern, all four types of GNNs need to consider graph topology. Additionally, SHeGNNs and DHeGNNs need to account for semantic information exposed by metapaths, while DHoGNNs and DHeGNNs must also consider temporal information due to the evolving graph.
In terms of execution bounds, these GNN models face hybrid execution bounds: memory-bound in graph-topology-based or RNN-based operations and compute-bound in MLP-based operations.

Based on this summary, we present a case study to illustrate how these findings are applied.

\textit{A Case Study.} We present a case study to demonstrate the practical application of these findings using HyGCN~\cite{HyGCN}, a hardware accelerator tailored for GNNs. 
First, informed by the finding that the gather and apply\_vertex stages dominate execution time (\RefTimeFinding{number:HyGCN:aggregation_combination_time}), specialized hardware units are designed to accelerate these critical stages. 
Second, leveraging the insight that GNNs experience hybrid execution bounds (\RefExecutionFinding{number:HyGCN:hybrid_execution_pattern}), an aggregation engine and a combination engine are introduced to address memory-bound and compute-bound challenges, respectively. 
Finally, based on the observation of inter-stage dataflow between these phases (\RefExecutionFinding{number:ieeecal2020_gcn_gpu:inter_stage_dataflow}), a pipeline is established between the two engines. This design leverages dataflow opportunities to reduce unnecessary off-chip memory accesses, thereby enhancing overall efficiency.

Based on this summary, we present several key challenges faced in GNN deployment.

\textit{Heterogeneity in Graph Types and Data Sources.} GNNs are applied across diverse domains, each with its own unique characteristics. For instance, social networks, biological networks, and knowledge graphs exhibit distinct graph structures and data distributions. The heterogeneity of these data sources makes it challenging to develop a one-size-fits-all deployment strategy. Adapting GNN models to different types of graphs involves customizing the model architecture, data preprocessing pipelines, and training procedures for each domain. Additionally, maintaining consistency, efficiency, and interoperability across these varied graph types and data sources presents a significant challenge in large-scale GNN deployment.

\textit{Graph Partitioning and Distribution.} Efficient graph partitioning is a critical challenge in deploying GNNs on distributed systems, especially when working with large graphs. The manner in which a graph is divided and distributed across multiple machines has a direct and substantial impact on the performance of the model. Poor partitioning and distribution can lead to increased communication overhead, network congestion, and higher latency due to frequent data exchanges between machines. This not only slows down performance but also reduces scalability.

\textit{Scalability with Large-Scale Graphs.} A key challenge in deploying GNNs is scaling them to efficiently handle very large graphs. These graphs can consist of billions of vertices and edges, and the compute complexity of GNNs increases significantly as the graph size grows. Scaling GNNs in a distributed environment requires careful attention to memory management, the design of parallelization strategies, and the use of distributed computing frameworks. 

\textit{Real-Time Processing.} Many GNN applications, such as recommendation systems, require real-time or low-latency inference. This necessitates not only fast data retrieval and processing but also the ability to deploy GNNs efficiently in environments where model updates or inferences must occur in real-time. Achieving real-time processing presents a significant challenge, as it demands high throughput and low-latency computation, which becomes particularly difficult when working with large graphs and complex models. 

\textit{Data Privacy and Security Concerns.} When dealing with sensitive data, such as personal information in social networks or medical data in healthcare, data privacy and security become crucial considerations. Ensuring that the distributed processing of graph data does not expose sensitive information is challenging. Techniques such as differential privacy and federated learning are often employed to mitigate privacy concerns, but they introduce additional computational overhead and complexity. Ensuring that sensitive graph data remains secure during processing while maintaining the utility of the model is an ongoing challenge.

\section{Future Directions and Conclusion} \label{sec:direction_conclusion}
In this section, we first introduce several future directions and then provide a conclusion.

\subsection{Future Directions} \label{sec:direction_conclusion:direction}

We introduce the future directions for GNN characterizations and analyze their necessity as follows.

\textit{Characterization on Performance and Scalability.}
A future direction lies in conducting systematic characterization to compare different types of GNN models, which is vital for optimizing their performance. This effort involves uncovering the shared traits and unique distinctions among various GNN models from diverse architecture perspectives. By systematically identifying and quantifying performance bottlenecks, this direction offers critical insights for designing targeted software and hardware optimizations. Moreover, it helps prevent the inadvertent propagation of inefficiencies from one model type to another. Ultimately, it paves the way for developing more refined and context-aware GNN architectures, ensuring they effectively address challenges specific to their type and deployment scenarios.

A promising direction in GNN characterization is the quantitative analysis of scalability, particularly in the context of large-scale graphs. As GNNs are applied to increasingly massive datasets, understanding how different models scale is essential for optimizing performance. This entails examining how GNNs manage graph size, vertex degree, and edge density while identifying key bottlenecks in data distribution, memory access, and parallelism. By systematically quantifying scalability, this direction provides critical insights for refining GNN architectures, enabling them to handle large graphs more effectively, scale efficiently across distributed systems, and maintain high performance under diverse computational demands.

A promising direction in GNN characterization is the quantitative analysis of graph partitioning techniques, which are pivotal for enhancing the scalability and efficiency of GNN models on large-scale graphs. This analysis examines partitioning methods with respect to graph topology and size, evaluating their impact on both performance and resource utilization. By integrating advanced graph partitioning techniques into the design of GNN architectures, we can achieve improved parallelization, reduced communication costs, and enhanced scalability, thereby optimizing GNN performance in large-scale distributed systems.

\textit{Characterization on Cross-domain Integration.} A promising direction in GNN characterization is the analysis and optimization of GNN architectures designed for multi-modal graph data, where graph structures integrate diverse formats such as text, images, and time-series data. By assessing the computational and memory requirements specific to multi-modal data, this approach identifies key bottlenecks in feature fusion and cross-modal interaction modeling. These insights facilitate the development of specialized GNN architectures that can efficiently process and integrate heterogeneous data types, thereby broadening their applicability in advanced domains such as multi-modal recommendation systems, autonomous systems, and cross-domain knowledge graphs.

\textit{Characterization on Auxiliary Aspects.} A key future direction in GNN characterization is the systematic examination of data privacy and security concerns, especially when handling sensitive information such as personal or medical data. This includes characterizing the impact of privacy-preserving techniques, such as differential privacy and federated learning, on GNN performance, particularly regarding computational overhead, scalability, and other related factors. Such studies will provide valuable insights into the trade-offs between ensuring data protection and maintaining model utility.

\textit{Characterization on Industrial Scenarios.} A crucial direction in GNN characterization is examining GNNs in industrial scenarios, such as recommendation systems, where thousands of tasks are executed for parallel real-time inference. By thoroughly understanding performance bottlenecks and resource utilization in these practical contexts, we can provide developers and engineers with valuable insights for optimization strategies. Additionally, recognizing these real-world constraints helps design GNN architectures that scale effectively while maintaining high performance in production environments. This characterization establishes a strong foundation for future advancements in graph-based machine learning technologies.

\textit{Automation and Tooling for Characterization.} A promising direction in GNN characterization is the development of fast, automated tools for performance analysis and exploration. These tools enable the identification and optimization of performance bottlenecks in GNNs through real-time profiling, predictive performance modeling, and automated tuning. They accelerate GNN architecture optimization by quickly identifying inefficiencies, enhancing the ability to fine-tune GNN architectures in real-time.

\subsection{Conclusion} \label{sec:direction_conclusion:conclusion}

Our survey aims to help scholars systematically understand GNN performance bottlenecks and patterns from a computer architecture perspective, contributing to more efficient GNN execution. By providing a detailed overview of the current state of GNN characterization, this work not only aids researchers in avoiding redundant efforts but also accelerates the development of optimized GNN implementations. This, in turn, supports the broader adoption and advancement of GNNs, ultimately pushing the boundaries of what these powerful models can achieve.



\ifCLASSOPTIONcompsoc
    \section*{Acknowledgments}
\else
    \section*{Acknowledgment}
\fi

This work was supported by National Key R\&D Program of China (Grant No. 2023YFB4503500), National Natural Science Foundation of China (Grant No. 62202451 and No. 6230247), CAS Project for Young Scientists in Basic Research (Grant No. YSBR-029), CAS Project for Youth Innovation Promotion Association, Beijing Natural Science Foundation (Grant No. L234078), and Beijing Nova Program (Grant No. 20220484054 and No. 20230484420).


\ifCLASSOPTIONcaptionsoff
  \newpage
\fi



%



\bibliographystyle{IEEEtranS}
\bibliography{refs}

%

\section{Biography Section}

\begin{IEEEbiography}[{\includegraphics[width=1in,height=1.25in,clip,keepaspectratio]{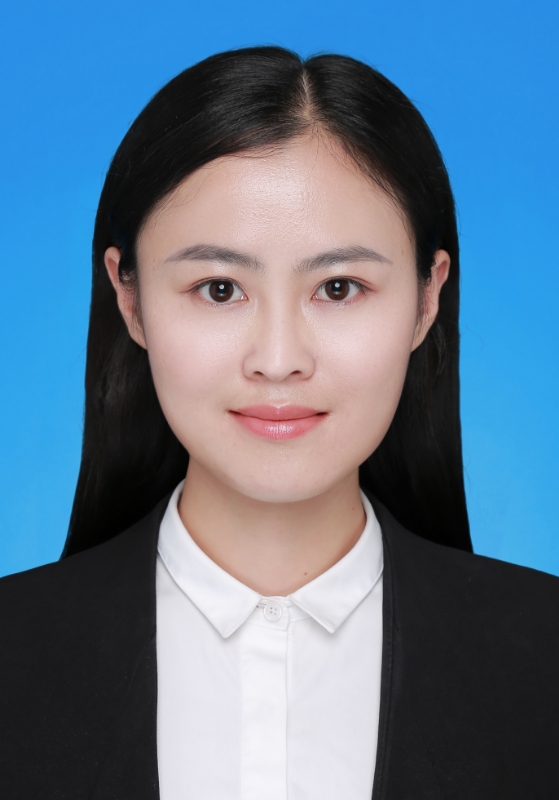}}]{Meng Wu} received the M.Eng. degree from University of Chinese Academy of Sciences, Beijing, China, in 2017. She is currently working toward the Ph.D. degree at Institute of Computing Technology, Chinese Academy of Sciences, Beijing, China. Her current research interests include graph-based hardware accelerator, high-throughput computer architecture, and parallel algorithm. To date, she has published over 10 research papers in compute journals and conferences, including IEEE TPDS, FCGS, IEEE CAL, J INF SCI, NOCS, Euro-Par and so on.
\end{IEEEbiography}

\begin{IEEEbiography}[{\includegraphics[width=1in,height=1.25in,clip,keepaspectratio]{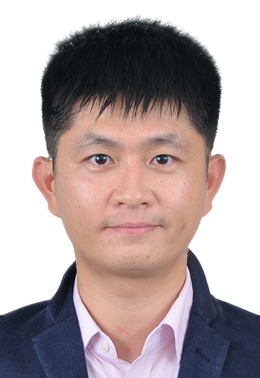}}]{Mingyu Yan} received the Ph.D. degree from University of Chinese Academy of Sciences, Beijing, China in 2020. He is currently an Associate Professor at Institute of Computing Technology, Chinese Academy of Sciences, Beijing, China. His current research interests is domain-specific hardware architecture for graph-based machine learning. To date, Dr. Yan has published over 20 research papers in top-tier journals and conferences, including the MICRO, HPCA, AAAI, IJCAI, DAC, ICCAD, PIEEE, IEEE TPDS, IEEE TC, IEEE TCAD, IEEE/CAA JAS, IEEE J-STSP, and so on. He has served as the TPC or ERC member for ISCA, HPCA, MICRO, and ICS.

\end{IEEEbiography}

\begin{IEEEbiography}[{\includegraphics[width=1in, height=1.25in, clip, keepaspectratio]{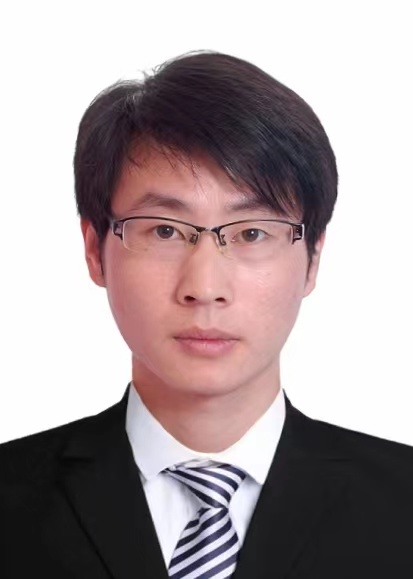}}]{Wenming Li}
received the Ph.D. degree in computer architecture from Institute of Computing Technology, Chinese Academy of Sciences, Beijing, in 2016. He is currently an associate professor in Institute of Computing Technology, Chinese Academy of Sciences, Beijing. His main research interests include high-throughput processor architecture, dataflow architecture and software simulation. To date, Dr. Li has published over 30 research papers in compute journals and conferences, including the HPCA, DAC, IEEE TC, IEEE TPDS, IEEE TACO, IEEE TCAD, and so on. 

\end{IEEEbiography}

\begin{IEEEbiography}[{\includegraphics[width=1in,height=1.25in,clip,keepaspectratio]{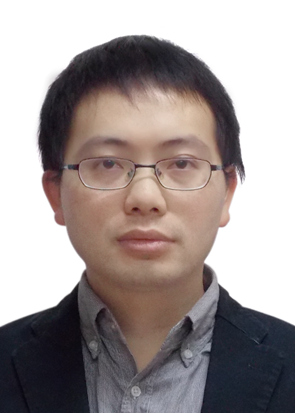}}]{Xiaochun Ye} received the Ph.D. degree in computer architecture from Institute of Computing Technology, Chinese Academy of Sciences, Beijing, in 2010. 
He is currently a Professor at Institute of Computing Technology, Chinese Academy of
Sciences, Beijing. 
His main research interest is domain-specific hardware architecture for graph-based machine learning and high-throughput computer architecture.
To date, Dr. Ye has published over 90 research papers in compute journals and conferences, the MICRO, HPCA, PACT, IPDPS, PIEEE, IEEE TC, IEEE TPDS, IEEE TACO, etc.

\end{IEEEbiography}

\begin{IEEEbiography}[{\includegraphics[width=1in,height=1.25in,clip,keepaspectratio]{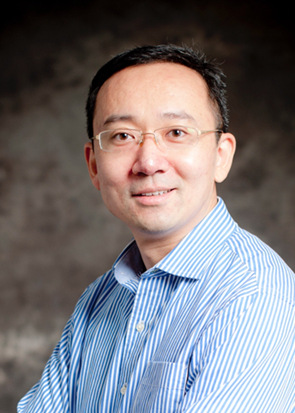}}]{Dongrui Fan} received the Ph.D. degree in computer architecture from Institute of Computing Technology, Chinese Academy of Sciences, Beijing, in 2005. 
He is currently a Professor and Ph.D. Supervisor at Institute of Computing Technology, Chinese Academy of Sciences, Beijing. 
His main research interests include high-throughput computer architecture, high-performance computer architecture, and low-power design.
To date, Dr. Fan has published over 140 research papers in compute journals and conferences, including the MICRO, HPCA, PPoPP, IJCAI, PACT, PIEEE, IEEE TC, IEEE TPDS, IEEE TCAD, IEEE TACO, IEEE Micro, and so on. He has been a member of the program committee of many important academic conferences, including MICRO, HPCA, etc.

\end{IEEEbiography}

\begin{IEEEbiography}[{\includegraphics[width=1in,height=1.25in,clip,keepaspectratio]{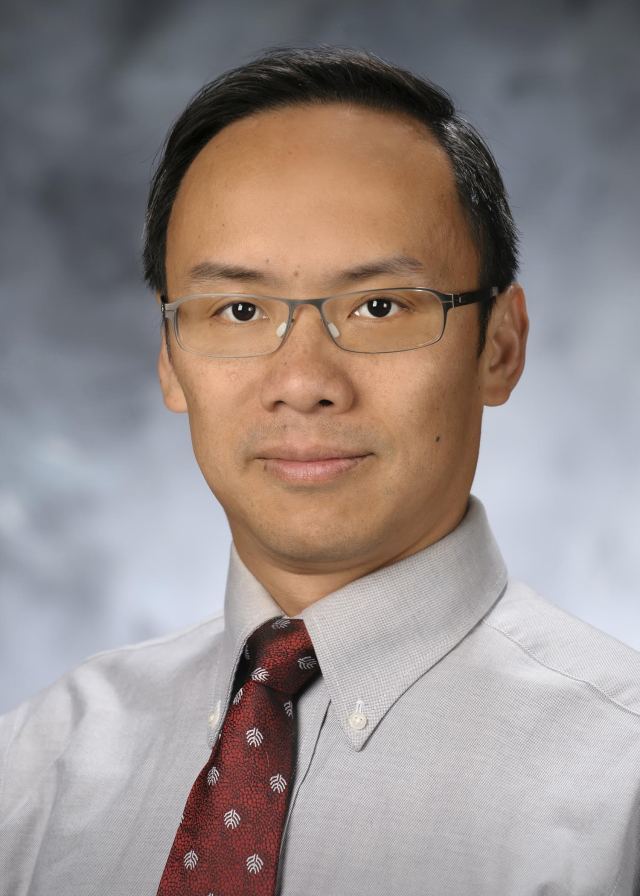}}]{Yuan Xie} (Fellow, IEEE) received the B.S. degree in electronic engineering from Tsinghua University, Beijing, China, in 1997, and the M.S. and Ph.D. degrees in electrical engineering from Princeton University, Princeton, NJ, USA, in 1999 and 2002, respectively. He has a rich industry experience with IBM, AMD, and Alibaba Group. He was also on the faculty of Pennsylvania State University and University of California, Santa Barbara. He is a Fellow of IEEE, a Fellow of ACM, and a Fellow of AAAS. He is a recipient of many awards, including NSF CAREER Award (2006), IEEE Computer Society Edward J. McCluskey Technical Achievement Award (2021), and IEEE CAS Society Industrial Pioneer Award (2023). 
He is currently a Professor with the Department of Electrical and Computer Engineering, Hong Kong University of Science and Technology, Hong Kong, SAR, China. 
His research interests include VLSI design, electronics design automation (EDA), computer architecture, and embedded systems.

\end{IEEEbiography}

\vfill









\end{document}